\documentclass[twocolumn,pre,eqsecnum,showpacs]{revtex4}
\usepackage{graphicx}

\begin{document}

\title[Hamiltonian walks on fractals]{Scaling of Hamiltonian walks on fractal lattices}

\author{Sun\v cica \surname{Elezovi\' c-Had\v zi\' c}}\email{suki@ff.bg.ac.yu}\affiliation{
University of Belgrade, Faculty of Physics, P.O. Box 368,
Belgrade, Serbia}
\author{Du\v sanka \surname{Mar\v ceti\' c}}\email{dusamar@netscape.net}\affiliation{
University of Banja Luka, Faculty of Science, Department of Physics,
M.~Stojanovi\' ca 2, Banja Luka, Bosnia and Herzegovina}
\author{Slobodan \surname{Maleti\' c}}\email{supersloba@vin.bg.ac.yu}\affiliation{ Institute of Nuclear Sciences Vin\v
ca, P.O.Box 522, Belgrade, Serbia}

\date{\today}

\begin{abstract}
We investigate asymptotical behavior of numbers of long Hamiltonian walks (HWs), {\em i.e.} self-avoiding random walks that visit every site of a lattice, on various fractal lattices. By applying an exact recursive technique we obtain scaling forms for open HWs on 3-simplex lattice, Sierpinski gasket, and their generalizations: Given-Mandelbrot (GM), modified Sierpinski gasket (MSG) and $n$-simplex fractal families. For GM, MSG and $n$-simplex lattices with odd values of $n$, the number of open HWs $Z_N$, for the lattice with $N\gg 1$ sites, varies as $\omega^N N^\gamma$. We explicitly calculate the exponent $\gamma$ for several members of GM and MSG families, as well as for $n$-simplices with $n=3,5$, and 7. For $n$-simplex fractals with even $n$ we find different scaling form: $Z_N\sim \omega^N \mu^{N^{1/d_f}}$, where $d_f$ is the fractal dimension of the lattice, which also differs from the formula expected for homogeneous lattices. We discuss possible implications of our results on studies of real compact polymers.
\end{abstract}


\maketitle

\section{Introduction}

Self--avoiding walks (SAWs) have long been used in the studies of configurational statistics of polymer chains in solution \cite{Vanderzande}. Due to excluded volume effect, at high temperatures $T$ (good solvent) long polymer chains are in swollen configurations. At low temperatures (poor solvent) polymers are in a collapsed state, caused by the attractive interactions of different sections of a polymer, mediated by a solvent. The transition between these two states occurs at the so--called $\theta$ temperature, at which excluded volume and attractive forces balance. Whereas the swollen and $\theta$ phases has been well investigated by now, the entropic scaling of the collapsed phase is still an open issue. A closely related problem is the scaling of Hamiltonian walks (HWs), which are SAWs that visit all the sites of the underlying lattice. HWs are believed to represent the $T=0$ limit of collapsed polymers, and they are also used in the studies of polymer melting \cite{JacobsenKondev}, as well as in the context of protein folding  \cite{Lua}.

The number $Z_N$ of HWs on homogeneous lattices with $N \gg 1$ sites is expected to behave as
\begin{equation}
Z_N\sim \omega^N {\mu^{N^\sigma}_S}N^a\, .
\label{eq:asimptotika}
\end{equation}
Here $\sigma=(d-1)/d$, where $d$ is the dimensionality of the lattice, $\mu_S$ is some constant less than 1, and $\omega$ is the connectivity constant, defined as
\begin{equation}
 \ln\omega  =\lim_{N\to\infty}{{\ln Z_{N}}\over  N}\, .
 \label{eq:definicija}
\end{equation}
Proposed scaling form for HWs differs
from the ordinary SAW case (swollen polymer), where average number of $N$-step SAWs, for large $N$, behaves as $\omega^N N^a$, and where the critical exponent $a$ depends only on $d$ (which is not the case for HWs). The term ${\mu^{N^\sigma}_S}$ in the HW case is expected on the basis of the exact study of HWs on the Manhattan lattice \cite{DuplantierDavid}, as well as on the conjecture that collapsed polymer (globule) has a sharp boundary, so that a surface tension term should arise \cite{Owczarek}.
The scaling form (\ref{eq:asimptotika}) was confirmed by Owczarek \cite{SamoOwczarek} for collapsed partially directed SAWs on the square lattice. Baiesi {\em et al} \cite{Baiesi}
recently performed extensive Monte Carlo simulations, which gave strong evidence that (\ref{eq:asimptotika}) is also satisfied for undirected collapsed SAWs on the square lattice. There are hardly any results for higher dimensional lattices. To the best of our knowledge, the only clear indication that for collapsed SAWs on three-dimensional lattices there exists a surface term, as predicted by (\ref{eq:asimptotika}), was obtained by Grassberger and Hegger \cite{Grassberger} via Monte Carlo simulations. There are also no results concerning scaling form for the collapsed SAWs on disordered lattices. Having all that in mind it might be useful to study HWs on fractals.

Fractal lattices are somehow intermediate between homogeneous and disordered ones, and
their hierarchical and scale invariant structure often allows an exact recursive treatment of various physical phenomena.  The ordinary SAW model (corresponding to the polymer chain at high temperatures) has been studied extensively in the past on different fractals  \cite{DharNSimplex}-\cite{Marini}. These studies contribute to a better understanding of how dimension and topological structure of underlying space affect the critical behavior of SAWs in general. There are fewer papers about self-interacting SAWs on fractals \cite{Klein}-\cite{Dragica}, and neither of them establishes the explicit form of the partition function at low temperatures. The closely related problem of finding the number of walks in the limiting HW case was analyzed only for closed walks on some fractals \cite{Bradley,Stajic}.
Here for the first time we investigate the scaling forms for the number of open HWs on a set of fractal lattices.

The paper is organized as follows.
In Sec.~\ref{sec:SG} we describe the technique used for an exact enumeration of closed HWs on the cases of 3-simplex lattice and Sierpinski gasket fractal, and extend it for enumeration of open HWs on these lattices. In Sections \ref{sec:GM} and  \ref{sec:MSG} this technique is generalized for enumeration of HWs on Given-Mandelbrot (GM) and modified Sierpinski gasket (MSG) families of fractals. We find that there is no surface term in the scaling form for the number of HWs on neither of these lattices. For the number of closed HWs on MSG fractals we show that
it varies as $\omega^N$, which is the same as in the case of  GM fractals \cite{Stajic}.
For open HWs we obtain scaling form $\omega^N N^\gamma$ for both fractal families, with $\omega$ and $\gamma$ depending on the particular parameters of the fractals. Furthermore, exponent $\gamma$ does not have the same value for corresponding fractals from these two families, as is the case for ordinary SAWs \cite{NeZnam,EKM}. Similar violation of the universality was found for two-dimensional (2d) homogeneous lattices \cite{jane}. We extend our analysis to higher dimensional $n$-simplex lattices in Sec.~\ref{sec:nsimplex}, where general exact scheme for obtaining numbers of HWs is described. It turns out that scaling form for HWs strongly depends on the parity of the fractal parameter $n$. In particular, for lattices with odd values of $n$ we find the same scaling form as for HWs on GM and MSG fractals, whereas for even $n$ we get $Z_N\sim \omega^N \mu^{N^{1/d_f}}$, with $d_f$ being the fractal dimension of the lattice, and different values of $\mu$ for closed and open HWs. All obtained results are summarized in Sec.~VI, and some technical details are given in the appendixes.

\section{Hamiltonian walks on 3-simplex lattice and Sierpinski gasket fractal}  \label{sec:SG}

In order to explain the method used in this paper for obtaining the
scaling forms of numbers of both closed and open HWs we start with
the simplest case: HWs on the 3-simplex lattice, and proceed with HWs on
the Sierpinski gasket.

\subsection{3-simplex lattice}

\begin{figure}
\includegraphics[width=70mm]{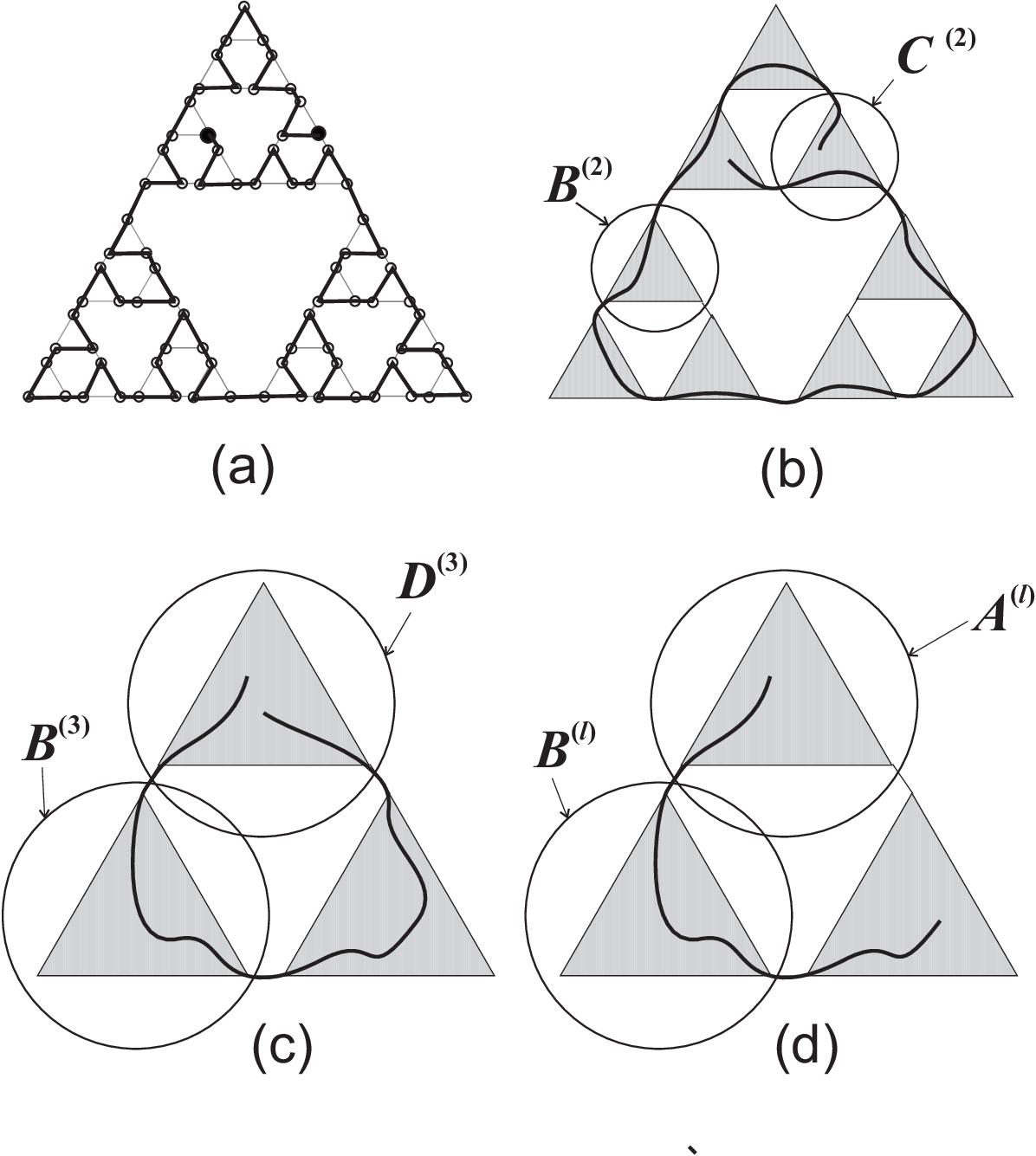}
\caption{ (a) An open HW on a 3-simplex of order $l=4$. The first
(b) and the second step (c) of the coarse-graining process. Gray
triangles in (b) and (c) represent 3-simplices of order $l=2$ and
$l=3$, respectively, whereas curved lines correspond to the
coarse-grained walk. Different types of configurations within the
third and the second order simplices are encircled. (d) Open HW on
3-simplex of order $(l+1)$, consisting of two $A$-type parts and
one $B$-type part within the corresponding simplices of order $l$
(gray triangles).} \label{fig:3simplexHW}
\end{figure}

To obtain the 3-simplex lattice \cite{NelsonFisher} one starts with
a complete graph of three points and replaces each of these points
by a new complete graph of three points. The subsequent stages are
constructed self-similarly, by repeating this procedure. After $l$
such iterations one obtains {\it 3-simplex of order $l$} (see
Fig.\ \ref{fig:3simplexHW}(a)), whereas the complete $3$-simplex
lattice is obtained in the limit $l\to\infty$. The number of
vertices within the $l$th order 3-simplex is $N_l=3^l$, and the
fractal dimension of the lattice is $d_f=\ln 3/\ln 2$.

In Fig.\ \ref{fig:3simplexHW}(a) we give an example of open HW on a
3-simplex of order $l=4$. Performing a coarse-graining process
 one notices in Fig.\ \ref{fig:3simplexHW}(b) that this walk can be
decomposed into parts corresponding to the second order simplices,
which have two possible configurations: one traversing the simplex
($B$-type configuration) and the other consisting of two strands,
one of which is traversing the simplex, and the other with one end at a corner
vertex of the same simplex, and the second end anywhere within it
($C$-type configuration). The coarse-graining can be applied once
more (Fig.\ \ref{fig:3simplexHW}(c)), leading to a HW consisting of
three parts corresponding to three simplices of order $l=3$, two
of them being of the type $B$, and the third consisting of two
strands (``legs''), each of them having one end-point at the
corner vertex of the corresponding simplex and the other within it
($D$-type configuration).
In a similar way, it is not difficult to see that each open HW on
the $(l+1)$th order simplex can be decomposed in two ways, either
as in the sample shown in Fig.\ \ref{fig:3simplexHW}(c), or as in
Fig.\ \ref{fig:3simplexHW}(d), with one $B$-type part and two
$A$-type parts (one strand with one end at the corner vertex and
the other end anywhere within the corresponding $l$th order
simplex). If we denote the number of open HWs on the $(l+1)$th
order simplex by $Z_O^{(l+1)}$,  then the following
relation is valid:
\begin{equation}
Z_O^{(l+1)}=3B^{(l)}\left[\left(A^{(l)}\right)^2+B^{(l)}D^{(l)}\right]\,
, \label{eq:otvorene}
\end{equation}
where $A^{(l)}$, $B^{(l)}$ and
$D^{(l)}$ are  the numbers of HWs of types $A$, $B$
and $D$ within the $l$-th order simplex, respectively.
Decomposition of closed HWs is even simpler, leading to the recursion
relation:
\begin{equation}
Z_C^{(l+1)}=\left(B^{(l)}\right)^3\, , \label{eq:zatvorene}
\end{equation}
where  $Z_C^{(l+1)}$ is the overall number of the closed HWs on the
simplex of order $(l+1)$.
\begin{figure}
\includegraphics[width=80mm]{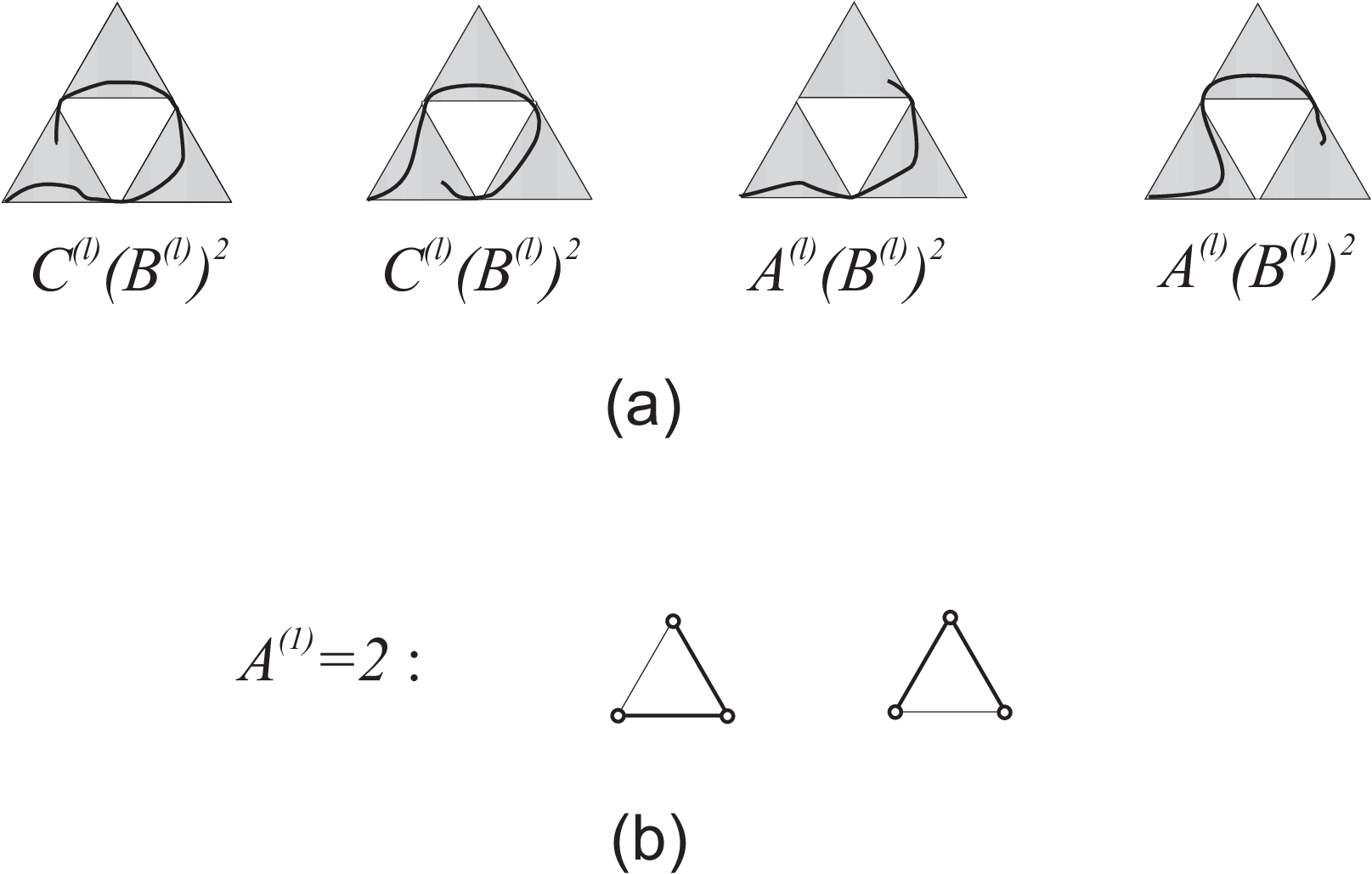}
\caption{(a) Four possible configurations of A-type HWs on a
3-simplex of order $(l+1)$, with the corresponding numbers of HWs.
Gray triangles represent the $l$th order simplex. (b) Two possible
A-type HWs on the first order 3-simplex. Small circles represent
sites the lattice consists of.} \label{fig:3simplexA}
\end{figure}
Number $C^{(l)}$ of $C$-type HWs  does not contribute directly to
$Z_O^{(l+1)}$ and $Z_C^{(l+1)}$, but one can obtain (see
Fig.\ ~\ref{fig:3simplexA}) the following recursion relations:
\begin{eqnarray}
 A^{(l+1)}&=&2B^2(A+C)\, ,\label{eq:a}\\
B^{(l+1)}&=&B^3 \, ,\label{eq:b}\\
C^{(l+1)}&=&B^2(A+3C)\, ,  \label{eq:c}\\
D^{(l+1)}&=&B(4AC+2A^2+3BD+6C^2)\, , \label{eq:d}
\end{eqnarray}
for the numbers of HWs corresponding to simplices of two
successive orders (where we have omitted index $(l)$ on the
right-hand side of the relations). Iterating these relations,
starting with the initial values $A^{(1)}=2$, $B^{(1)}=1$,
$C^{(1)}=1$, and $D^{(1)}=2$, one can calculate $A^{(l)}$,
$B^{(l)}$, $C^{(l)}$, and $D^{(l)}$ for any $l$, and,
consequently, the numbers $Z_O^{(l+1)}$ and $Z_C^{(l+1)}$.
Furthermore, these relations can be exactly analyzed, as follows.
From (\ref{eq:b}), and the corresponding initial condition,
trivially follows $B^{(l)}\equiv 1$, which also transforms the set
of relations (\ref{eq:a}) i (\ref{eq:c}) for one-leg walks $A$ and
$C$ into a simple system of difference equations:
\[
A^{(l+1)}=2(A^{(l)}+C^{(l)})\, ,  \quad C^{(l+1)}=A^{(l)}+3C^{(l)}\,
, \] whose solution is
\begin{equation}
A^{(l)}={1\over 3}(4^l+2)\, , \quad C^{(l)}={1\over 3}(4^l-1)\,
.\label{eq:ackonacno}
\end{equation}
Then, from recursion relation (\ref{eq:d}) straightforwardly follows
the difference equation for the number of two-leg walks $D$:
\[ D^{(l+1)}={2\over 3}4^{2r}+3D^{(l)}\, ,\] whose solution is
\begin{equation}
D^{(l)}={4\over{39}}16^r+{3\over{13}}3^r-{1\over 3}\, .
\label{eq:dkonacno}
\end{equation}
Finally, putting (\ref{eq:ackonacno}), (\ref{eq:dkonacno}), and
$B^{(l)}=1$ in formulas for the overall numbers of open
(\ref{eq:otvorene}) and closed (\ref{eq:zatvorene}) HWs one gets
\[
Z_O^{(l)}={{25}\over{16\cdot 39}}16^l+{1\over
3}4^l+{3\over{13}}3^l+{1\over 3}\, , \quad Z_C^{(l)}=1\, .
\]
From the last equation trivially follows that connectivity constant
is $\omega=1$ and $Z_C^{(l)}=\omega^{N_l}$ for every $l$, whereas
the number of open HWs for $l\gg 1$ behaves according to the
asymptotic formula
\begin{equation}
Z_O^{(l)}\sim N_l^{{\ln 16}\over{\ln 3}}=\omega^{N_l}
N_l^{\gamma}\, , \label{eq:3simplexOpenAsimp}
\end{equation}
where $N_l=3^l$ is the number of vertices of the 3-simplex of order $l$, and $\gamma=\ln 16/\ln 3=2.52372\ldots$..

\subsection{Sierpinski gasket}

\begin{figure}
\includegraphics[width=70mm]{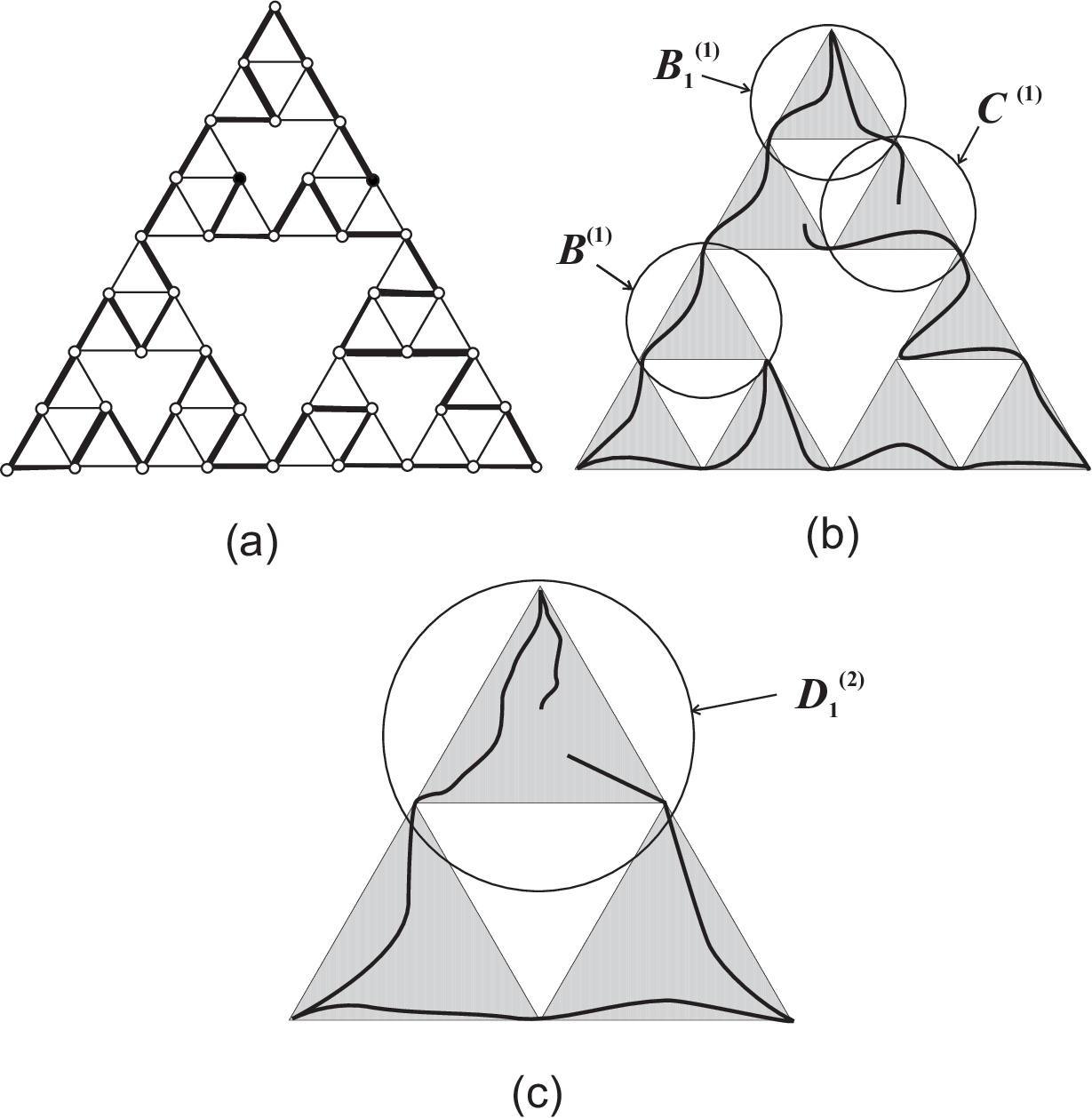}
\caption{ (a) An open Hamiltonian walk on Sierpinski gasket of
order $r=3$. Vertices corresponding to the end-points of the walk
are colored black, for the sake of easier recognition. The first
(b) and the second step (c) of coarse-graining process. Gray
triangles in (b) and (c) represent gaskets of order $r=1$ and
$r=2$, respectively, whereas curved heavy lines correspond to the
coarse-grained walk. Different kinds of configurations within the
first and the second order gaskets are encircled.}
\label{fig:SGb2HW}
\end{figure}

\begin{figure}
\includegraphics[width=50mm]{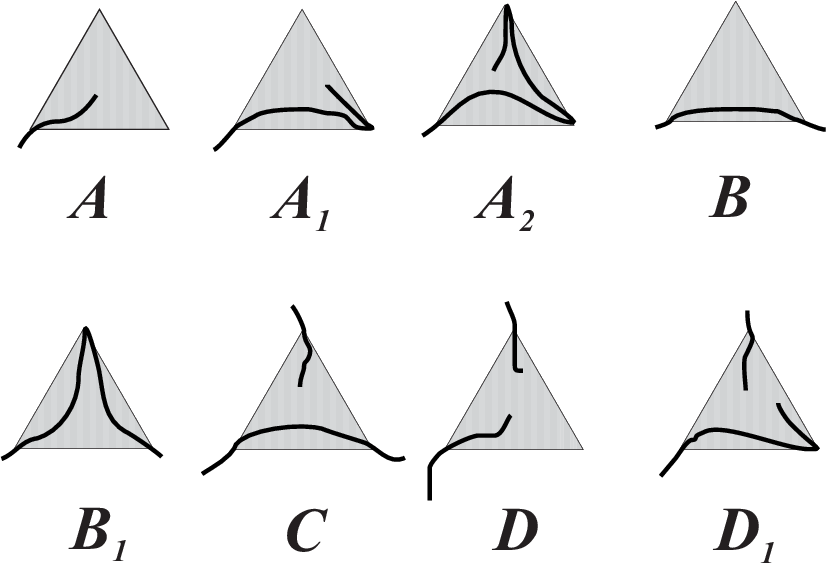}
\caption{Possible types of HWs on $r$th order gaskets (gray
triangles), needed to decompose any HW on gasket of order $r+1$.}
\label{fig:SGparametri}
\end{figure}

In a similar way, one can analyze HWs on the Sierpinski gasket
(SG) lattice. SG is a well known fractal lattice, which can be
constructed recursively, starting with the generator (gasket of
order $r=1$), which consists of three unit equilateral triangles,
arranged to form a twice larger triangle (see
Fig.\ ~\ref{fig:SGb2HW}(a)). The subsequent fractal stages are
constructed self-similarly, by replacing each of the unit
triangles of the initial generator with a new generator. To obtain
the $r$th–-stage fractal lattice ($r$th order gasket), this
process of construction has to be repeated $(r-1)$ times, and the
complete fractal is obtained in the limit $r\to\infty$. The
numbers of sites on the $r$th order gasket is equal to
$N_r={3\over 2}(3^r+1)$. SG resembles 3-simplex lattice and indeed
has the same fractal dimension $d_f=\ln 3/\ln 2$.
An open HW on a third order gasket is shown in
Fig.\ ~\ref{fig:SGb2HW}, together with its coarse-grained versions.
Comparing with Fig.\ ~\ref{fig:3simplexHW} one can observe that
larger number of types of possible HW configurations exists on SG
than in the case of 3-simplex lattice. There are exactly 8
different types of walks, and they are depicted in
Fig.\ ~\ref{fig:SGparametri}. The corresponding numbers of HWs on
the $r$th-order gasket will be denoted by $A^{(r)}$, $A_1^{(r)}$,
$A_2^{(r)}$, $B^{(r)}$, $B_1^{(r)}$, $C^{(r)}$, $D^{(r)}$, and
$D_1^{(r)}$. These numbers fulfill the following recursion
relations:
\begin{eqnarray}A^{(r+1)}&=&2\left(B^2C+ABB_1+
A_1B^2\right)\,, \label{eq:SGA}\\
A_1^{(r+1)}&=&A{B_1}^2+2A_1BB_1+2A_2B^2+2BB_1C\,,\label{eq:SGA1}\\
A_2^{(r+1)}&=&A_1{B_1}^2+2A_2BB_1+{B_1}^2C\,,\label{eq:SGA2}\\
B^{(r+1)}&=&2B^2B_1\,,\qquad B_1^{(r+1)}=2B{B^{2}_1}\,,\label{eq:SGBB1}\\
C^{(r+1)}&=&A{B_1}^2+2B(A_1B_1+A_2B+3B_1C),\label{eq:SGC}\\
D^{(r+1)}&=&4AB_1C+4ABA_2+4A_1BC+4BB_1D\nonumber\\
&+&4AB_1A_1+2B(3C^2+2BD_1+A_1^2),\label{eq:SGD}\\
D_1^{(r+1)}&=&4A_1A_2B+2AA_2B_1+2A_1CB_1
+4BB_1D_1\nonumber\\
&+&4A_2CB+3{C}^2B_1+{A_1}^2B_1+{B_1}^2D, \label{eq:SGD1}
\end{eqnarray}
with the initial values: $A^{(1)}=4$, $A_1^{(1)}=6$,
$A_2^{(1)}=3$, $B^{(1)}=3$, $B_1^{(1)}=2$, $C^{(1)}=4$,
$D^{(1)}=6$, and  $D_1^{(1)}=5$.
Iterating recursion relations (\ref{eq:SGA})-(\ref{eq:SGD1}) one can
calculate numbers of all possible types of HWs, in principle for
any $r$, and then eventually find the overall number $Z_O^{(r+1)}$ of open
 HWs on the $(r+1)$th order gasket, by putting them
into the formula:
\begin{equation}
Z_O^{(r+1)}=12\,{A_1}^{(r)}\,{A_2}^{(r)}\,{B_1}^{(r)} +
6\,({{B_1}^{(r)}})^2\,{D_1}^{(r)}\, .\label{eq:SGopen}
\end{equation}
However, this is a tedious task, since all of these numbers quickly become extremely large. To avoid that, we analyze recursion relations in a similar manner as was done for 3-simplex lattice.

Introducing new variables:
$x_r=A^{(r)}/B^{(r)}$, ${x_1}_r=A_1^{(r)}/B^{(r)}$,
${x_2}_r=A_2^{(r)}/B^{(r)}$, $y_r=C^{(r)}/B^{(r)}$,
$z_r=D^{(r)}/B^{(r)}$, ${z_1}_r=D_1^{(r)}/B^{(r)}$, and noticing
that $B_1^{(r)}/B^{(r)}=2/3$ is exactly satisfied for any $r$,
from (\ref{eq:SGA})-(\ref{eq:SGD1}) one gets the new recursion
relations:
\begin{eqnarray}
\left(
  \begin{array}{c}
    x' \\
    {x_1}' \\
    {x_2}' \\
    y' \\
  \end{array}
\right) &=&\left(\matrix{ 1 & \frac{3}{2} & 0 & \frac{3}{2} \cr
\frac{1}{3} & 1 & \frac{3}{2} & 1 \cr 0 & \frac{1}{3} & 1 & \frac{1}
   {3} \cr \frac{1}{3} & 1 & \frac{3}{2} & 3 \cr  }\right)\left(
                                                            \begin{array}{c}
                                                              x \\
                                                              {x_1} \\
                                                              {x_2} \\
                                                              y \\
                                                            \end{array}
                                                          \right)\,
                                                          ,
                                                          \label{eq:SGxx1x2y}\\
\left(  \begin{array}{c}
    z' \\
    z_1' \\
  \end{array}
\right)&=&\left(
  \begin{array}{cc}
    2 & 3 \\
    {1\over 3} & 2 \\
  \end{array}
\right)\left(
         \begin{array}{c}
           z \\
           z_1 \\
         \end{array}
       \right)+\left(
                 \begin{array}{c}
                   f_1 \\
                   f_2 \\
                 \end{array}
               \right)\,
  ,\label{eq:SGz1}
\end{eqnarray}
where \begin{eqnarray}f_1&=&2x x_1+
\frac{3}{2}{{x^2_1}} + 3 x x_2 + 2x y +
  3 x_1 y+\frac{9}{2}y^2\, ,\nonumber\\
f_2&=& \frac{1}{2}{x^2_1} + x{x_2} + 3{x_1}{x_2} +
x_1y +
  3x_2 y + \frac{3}{2}\,y^2\, .\nonumber\end{eqnarray}
 From (\ref{eq:SGxx1x2y}) for $r\gg 1$
follows $x_r$, ${x_1}_r$,  ${x_2}_r$,  $y_r \sim \lambda^r$, where
$\lambda=4$ is the only eigenvalue of the $4\times 4$ matrix of this recursion relation which is larger than 1, and
then, from (\ref{eq:SGz1}), we find $z_r, {z_1}_r\sim 4^{2r}$.

\begin{figure}
\includegraphics[width=70mm,angle=270]{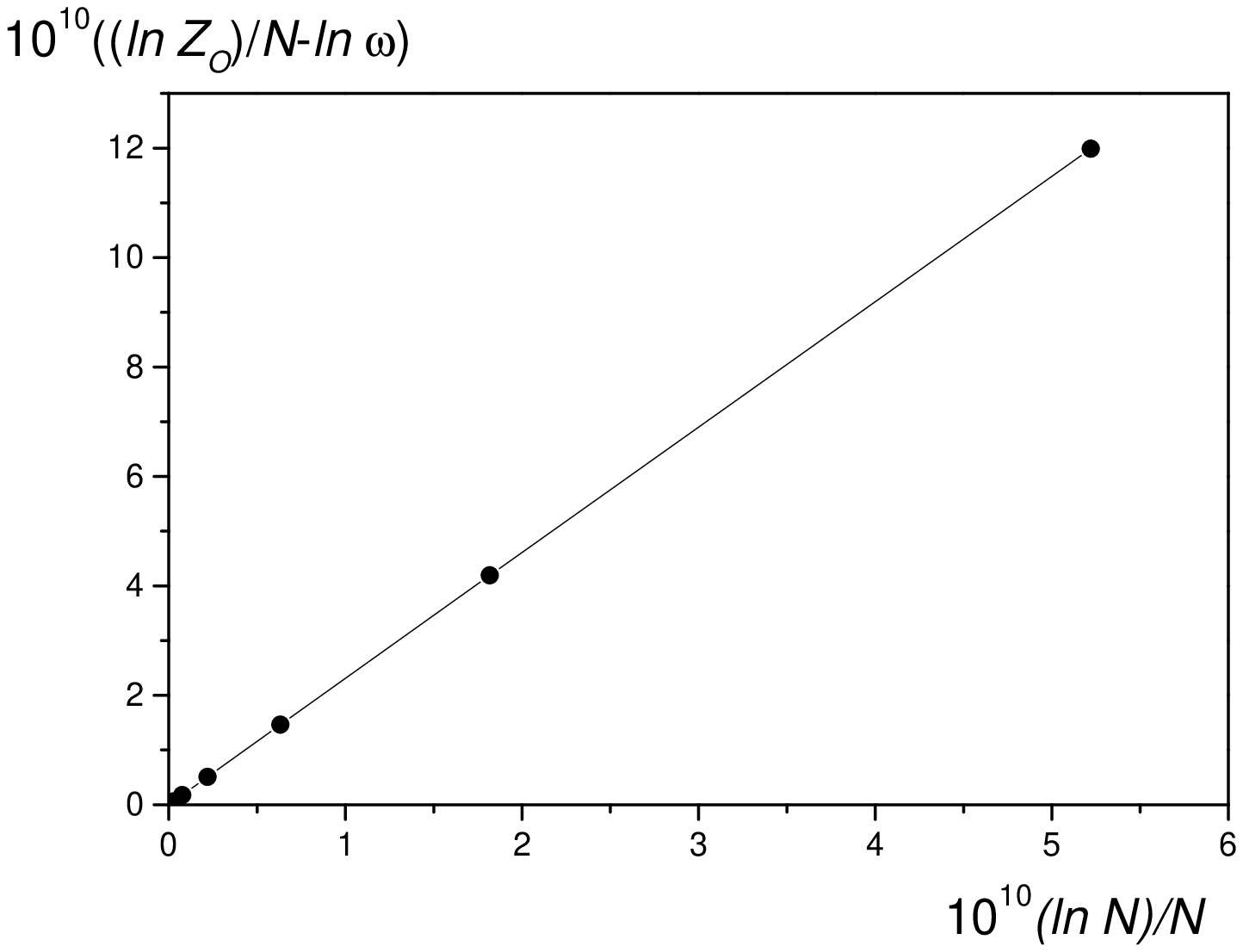}
\caption{Values of $(\ln Z_O^{(r)})/N_r$, found by direct numerical iteration of relations (\ref{eq:SGA})-(\ref{eq:SGD1}) and (\ref{eq:SGopen}), for $21\leq r\leq 30$. Line connecting the points has been obtained by linear fitting, and $\omega=12^{1/9}$.}
\label{fig:numerika}
\end{figure}

To obtain the asymptotic formula for the number of open
HWs,  we express $Z_O^{(r+1)}$ from (\ref{eq:SGopen})  as
\[
Z_O^{(r+1)}={8\over
3}\left(B^{(r)}\right)^3(3{x_1}_r{x_2}_r+{z_1}_r)\, .
\]
Since in \cite{Stajic} it was exactly shown  that
$B^{(r)}=\mathrm{const}\,\omega^{N_r}$, with $\omega=12^{1/9}$, one finally
gets the scaling form: $Z_O^{(r)}\sim\omega^{N_r}N_r^{\gamma}$, with $\gamma=\ln 16/\ln 3=2.52372\ldots$. For the sake of comparison, we have generated $Z_O^{(r)}$ by direct numerical iteration, for $r$ up to 100. In Fig.~{\ref{fig:numerika}} we depict $(\ln Z_O^{(r)})/N_r$, as a function of $(\ln N_r)/N_r$, for $21\leq r\leq 30$. As one can see, value of $(\ln Z_O^{(r)})/N_r$ for this range of $r$ is quite close to $\ln\omega$, whereas by fitting these data to linear function one obtains approximate value: $\gamma_{fit}=2.3$. For larger number of iterations $\gamma_{fit}$ becomes closer to the exact value of $\gamma$. For instance, fitting the numerical data for $91\leq r\leq 100$ gives $(\gamma-\gamma_{fit})/\gamma=2\%$.

Scaling form obtained for the number of open HWs on SG is the same as the one found for the 3-simplex lattice. As was shown in \cite{Stajic}, the numbers of closed HWs on SG scale as $\omega^{N_r}$, again the same as in the case of 3-simplex. Equality of the exponents for SG and 3-simplex is in accord with the fact that SAWs on these two lattices belong to the same universality class \cite{NeZnam}. On the other hand, it is known that exponents for HWs on different 2d Euclidean lattices have different values \cite{jane}, which is explained to be a consequence of the frustration, induced by the strong constraint that all the sites must be visited. It is believed that a relevant physical measure of this frustration is the number of contacts per monomer, {\em i.e.} vertex pairs which are not adjacent along the HW, but are the nearest neighbors on the lattice. Nevertheless, the number of contacts on 3-simplex is one, whereas it is two on SG, so that one could have expected different values of $\gamma$. In order to gain a deeper insight into the problem of universality and frustration of HWs on lattices embedded in 2d space, in the next two sections we analyze the asymptotic behavior of HWs on the appropriate generalizations of 3-simplex and SG fractals.

\section{Hamiltonian walks on Given-Mandelbrot fractals}\label{sec:GM}

One possible way to generalize the SG fractal is to start with a
generator that consists of $b(b+1)/2$ unit equilateral triangles,
arranged to form a $b$ times larger triangle. Enlarging the
generator $b$ times and substituting the smallest triangles with
the generator, and then repeating this procedure recursively {\em
ad infinitum}, one obtains fractal lattice characterized with the
integer $b$. For $b=2,3,...,\infty$ the complete so called
Given-Mandelbrot (GM) family of fractals is obtained
\cite{GivenMandelbrot}. SG is the first member of
this family, with the scaling parameter $b=2$. The fractal
dimension of GM fractal with scaling parameter $b$ is equal to
$d_f=\ln[b(b + 1)/2]/\ln b$, and the number of sites $N_r$ at the
$r$th stage of fractal construction is
\begin{equation}
N_{r}={{b+4}\over{b+2}}\left[{{b(b+1)}\over
2}\right]^{r}+2{{b+1}\over{b+2}}\, . \label{eq:GMcvorovi}
\end{equation}

The overall number of open HWs on the
$(r+1)$th stage of construction of any GM fractal can be expressed
in terms of numbers of 8 HW types within the $r$th order
stage, in a similar manner as in the case of SG (Fig.\ ~\ref{fig:SGparametri}).
Recursion relations for numbers of $B$, and $B_1$-type walks on two successive stages
of fractal construction have the following form:
\begin{equation}
B'=p\, B^{\frac{b(b-1)}{2}+1}\, {B_1}^{b-1}\, ,\quad
 B_1'=p\, {B}^{\frac{b(b-1)}{2}}\, {B_1}^{b}\,
 , \label{eq:bezkrake}
\end{equation}
as was shown in \cite{Stajic},
whereas the numbers of one-leg configurations: $A$, $A_1$, $A_2$, and $C$, satisfy a closed set
of recursion relations, which can be put in matrix form as:
\begin{equation}
\left(\begin{array}{c}
  A' \cr
  A_1' \cr
 A_2' \cr
  C'
\end{array}\right)=
\left(\matrix{ a_{11} & a_{12} & a_{13} &a_{14}\cr a_{21} & a_{22}
& a_{23} &a_{24}\cr a_{31}& a_{32} & a_{33} &a_{34}\cr a_{41} &
a_{42} & a_{43} &a_{44}  }\right)
  \left(\begin{array}{c}
  A \cr
  A_1 \cr
  A_2 \cr
  C
\end{array}\right)\, ,  \label{eq:GMmatricno}
\end{equation}
where $a_{ij}$ are polynomials in $B^{(r)}$ and $B_1^{(r)}$.
In Appendix~A we prove that each of these polynomials has only
one term, which is of the form $K_{ij}[B^{(r)}]^\beta
[B_1^{(r)}]^\delta$, where
$\beta+\delta=b(b+1)/2-1$. The coefficients $K_{ij}$ depend only
on $b$, and each of them can be expressed in terms of the number
$p$ (appearing in equations (\ref{eq:bezkrake})), and additional 8
numbers: $a, a_1, a_2, c, a', a_1', a_2', c'$. Introducing
new variables: $x_r=A^{(r)}/B^{(r)}$, ${x_1}_r=A_1^{(r)}/B^{(r)}$,
${x_2}_r=A_2^{(r)}/B^{(r)}$, $y_r=C^{(r)}/B^{(r)}$, relation
(\ref{eq:GMmatricno}) gives the new recursive relation:
\begin{equation}
\left(
  \begin{array}{c}
    x' \\
    {x_1}'\\
    {x_2}' \\
    y' \\
  \end{array}
\right) =\left(\matrix{ {a\over p} & \frac{a_1}{t\, p} &
\frac{a_2}{t^2\, p} & \frac{c}{t\, p} \cr \frac{t(2a-p)}{2\,p} &
\frac{a_1}{p} & \frac{a_2+p}{t\,p} & {c\over p} \cr
 \frac{t^2(a-p)}{2\,p} & \frac{a_1\,t}{2\,p}
& \frac{a_2+2p}{2\,p} & \frac{c\,t}{2\,p} \cr
 \frac{a't}{p} & \frac{a_1'}{p} &
\frac{a_2'}{t\,p} & {{c'}\over{p}} \cr  }\right)\left(
                                                            \begin{array}{c}
                                                              x\\
                                                              {x_1} \\
                                                              {x_2} \\
                                                              y \\
                                                            \end{array}
                                                          \right)\,
                                                          ,
                                                          \label{eq:GMxx1x2y}
\end{equation}
where $t=B_1^{(r)}/B^{(r)}$ is a coefficient that depends only on
$b$, as can be seen from (\ref{eq:bezkrake}). The particular
values of the coefficients appearing in the last 4x4 matrix, for $2\leq
b\leq 7$, are given in Appendix A.  For all $b$ considered here,
this matrix has only one eigenvalue $\lambda$ larger than 1,
implying that for $r\gg 1$ parameters $x_r$, ${x_1}_r$, ${x_2}_r$,
and $y_r$ grow as $\lambda^r$, and consequently, numbers
$A^{(r)}$, ${A_1}^{(r)}$, ${A_2}^{(r)}$, and $C^{(r)}$
asymptotically behave as $\lambda^r\, B^{(r)}$.

Numbers of two-leg HWs satisfy recursion relation:
\begin{equation}\left(
                  \begin{array}{c}
                    D' \\
                    {D_1}' \\
                  \end{array}
                \right)=\left(
                                  \begin{array}{cc}
                                    b_{11} & b_{12} \\
                                    b_{21} & b_{22} \\
                                  \end{array}
                                \right)\left(
                  \begin{array}{c}
                    D \\
                    {D_1}\\
                  \end{array}
                \right)+\left(
                          \begin{array}{c}
                            F \\
                            F_1 \\
                          \end{array}
                        \right)\, , \label{eq:GMTwoLeg}
\end{equation}
where $b_{ij}$ are polynomials in $B^{(r)}$ and $B_1^{(r)}$ of the
power $b(b+1)/2-1$ (see Appendix B), whereas $F$ and $F_1$ are of
the form:
$d_{AA}A^2+d_{AA_1}A\,A_1+d_{AA_2}A\,A_2+d_{AC}A\,C+d_{A_1A_1}A_1^2+d_{A_1A_2}A_1A_2+d_{A_1C}A_1C+
d_{A_2A_2}A_2^2+d_{A_2C}A_2C+d_{CC}C^2$, with $d_{XY}$ being
polynomials in $B^{(r)}$ and $B_1^{(r)}$ of the power
$b(b+1)/2-2$. With new parameters defined as $z_r=D^{(r)}/B^{(r)}$
and ${z_1}_r={D_1}^{(r)}/B^{(r)}$, from (\ref{eq:GMTwoLeg})
follows simpler recursion relation:
\begin{equation}\left(
                  \begin{array}{c}
                    z' \\
                    {z_1}' \\
                  \end{array}
                \right)=\left(
                                  \begin{array}{cc}
                                    m_{11} & m_{12} \\
                                    m_{21} & m_{22} \\
                                  \end{array}
                                \right)\left(
                  \begin{array}{c}
                    z \\
                    {z_1} \\
                  \end{array}
                \right)+\left(
                          \begin{array}{c}
                            f \\
                            f_1 \\
                          \end{array}
                        \right)\, , \label{eq:GMTwoLegLinearized}
\end{equation}
where $m_{ij}$ are numbers, which do not depend on $r$, but only
on $b$, whereas $f$ and $f_1$ are polynomials in $x_r$, ${x_1}_r$,
${x_2}_r$ and $y_r$. In particular, it can be shown  that
$m_{11}=m+1$, $m_{12}=2n/t$, $m_{21}=m t/2$, $m_{22}=n+1$, with
$m=b(b-1)/2$, and $n=b-1$, whereas $f$ and $f_1$ have the form:
$a_{xx}x^2+a_{xx_1}xx_1+a_{xx_2}xx_2+a_{xy}xy+a_{x_1x_1}x_1^2+a_{x_1x_2}x_1x_2+a_{x_1y}x_1y
+a_{x_2x_2}x_2^2+a_{x_2y}x_2y+a_{yy}y^2$, where $a_{XY}$ are again
numbers depending only on the scaling parameter $b$. It turns out that
for all $b$ considered here $\lambda^2$ is larger than eigenvalues
$\lambda_D$ of the matrix $m_{ij}$, implying that $z_r$,
${z_1}_r\sim \lambda^{2r}$.

The number $Z_O^{(r+1)}$ of open HWs  on the gasket of order
$(r+1)$, can be expressed as:
\begin{widetext}
\begin{eqnarray}
Z_O^{(r+1)}&=&k_1AA_{1}B^{m-3}B_{1}^{n+2}+k_2AA_{2}B^{m-2}B_{1}^{n+1}+k_3A_{1}A_{2}
B^{m-1}B_{1}^{n} +k_4ACB^{m-3}B_{1}^{n+2}\nonumber\\
&+&k_5A_{1}CB^{m-2}B_{1}^{n+1}+
k_6A_{2}CB^{m-1}B_{1}^{n}+k_7A^2B^{m-4}B_{1}^{n+3}+k_8A_{1}^2B^{m-2}B_{1}^{n+1}\nonumber\\
&+&k_9A_{2}^2B^{m}B_{1}^{n-1}+k_{10}C^2B^{m-2}B_{1}^{n+1}+k_{11}DB^{m-2}B_{1}^{n+2}+k_{12}D_{1}B^{m-1}B_{1}^{n+1}\,,
\label{eq:GMOpen}
\end{eqnarray}
\end{widetext}
where we have suppressed index $r$ on the right-hand side of this
relation, and $k_i$ are numbers that depend only on $b$. Substituting established asymptotical behavior of
$A_i$, $C$ and $D_i$ in the latter expression, one finds that all terms on the right-hand
side of equation (\ref{eq:GMOpen}) have the same asymptotical
form, so that
\[
Z_O^{(r+1)}\sim\lambda^{2r}(B^{(r)})^{\frac{b(b+1)}{2}}\, .
\]
Since in \cite{Stajic} it was shown that numbers $B^{(r)}$ and
$B_1^{(r)}$ for large $r$ behave as $\omega^{N_r}$, where the
number of sites $N_r$ is given by (\ref{eq:GMcvorovi}), it follows
that
\begin{equation}
Z_O^{(r)}\sim\omega^{N_r}N_r^{\gamma}\,, \quad  \mathrm{with}
\quad \gamma={2}\frac{\ln\lambda(b)}{\ln\frac{b(b+1)}{2}}\,
.\label{eq:openGM}
\end{equation}
The values of $\gamma$ for $2\leq b\leq 7$ are equal to $2.5237\ldots$,
$2.1841\ldots$, $2.3411\ldots$, $2.2461\ldots$, $2.2981\ldots$, and $2.2755\ldots$,
respectively. One should mention here that number of
closed HWs asymptotically behaves as $Z_C^{(r)}\sim \omega^{N_r}$,
as was established in \cite{Stajic}, where also a closed formula for $\omega$ was derived.

\section{Modified Sierpinski gasket fractals}\label{sec:MSG}

\begin{figure}
\includegraphics[width=70mm]{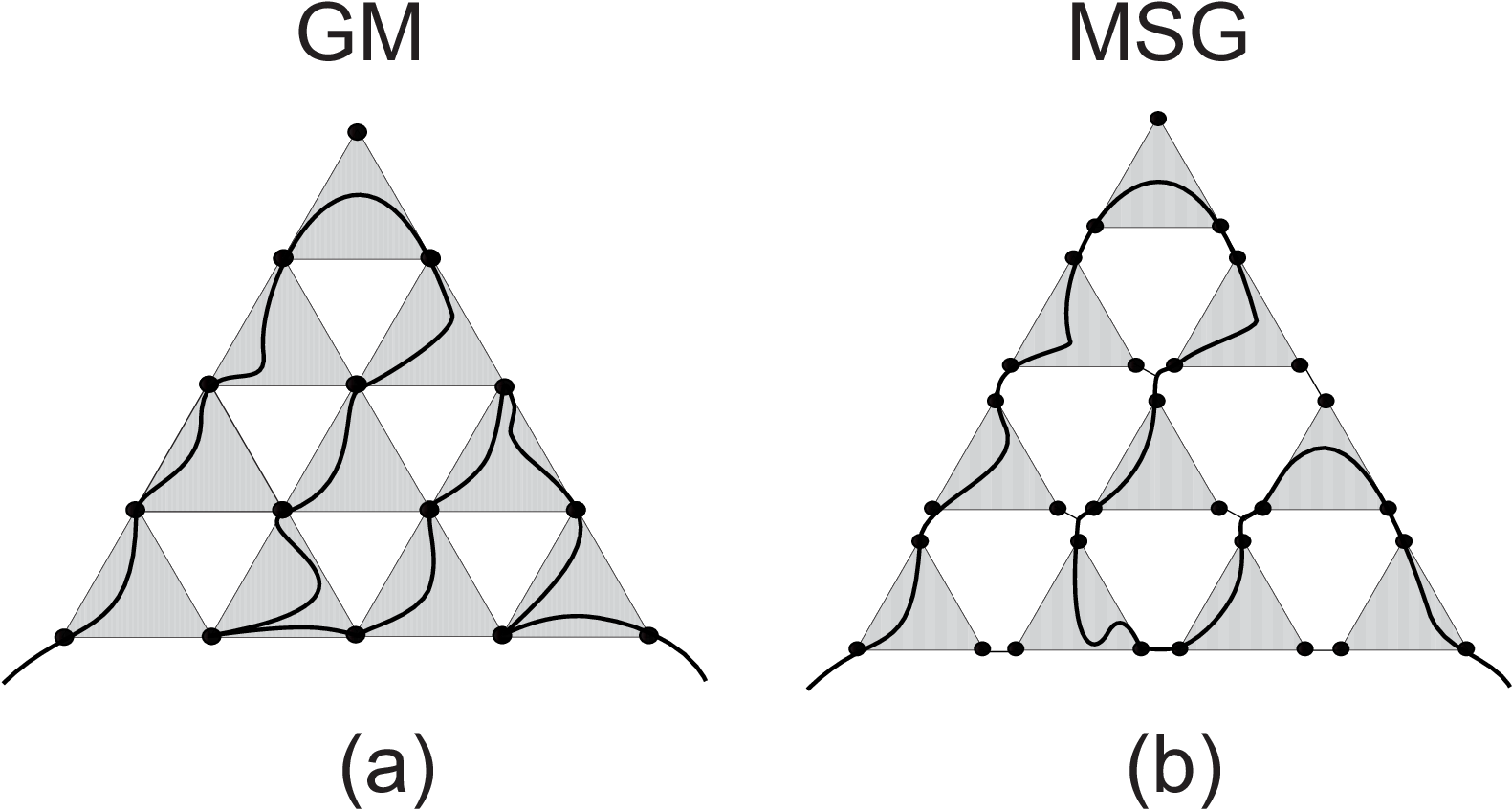}
\caption{ (a) The $(r+1)$th order generator of the GM fractal with
$b=4$. Grey triangles represent its 10 constitutive elements, each
of them being the $r$th stage of construction of the fractal
($r$th order generator). Most of the corner vertices (black circles) of the $r$th order generators belong to more than one triangle. (b) The $(r+1)$th order generator of the
MSG fractal with $b=4$ is obtained by connecting $r$th order
generators (grey triangles) via infinitesimal junctions. Curved
lines represent two similar $B$-type HWs which in (a) case
consists of $B$ and $B_1$ type strands within the $r$th order
triangles, whereas in (b) case each strand within the $r$th order
triangle is of $B$ type, implying that all vertices within the
triangle are visited (including the third corner vertex).}
 \label{fig:MSG}
\end{figure}

Intrigued by the fact that scaling relations for the numbers of HWs, as well as the corresponding values of the critical exponents, are the same for 3-simplex and SG lattices (Sec.~II), here we examine HWs on modified Sierpinski gasket (MSG) fractals.
In particular, the only difference between GM fractal and corresponding MSG fractal (with the same value of scaling parameter $b$) is that the generator of the MSG fractal is
obtained by arranging $b(b+1)/2$ unit triangles in such a way that
vertices of neighboring triangles are connected with an
infinitesimal junction, {\em i.e.} not glued, as in the GM case (see
Fig.\ ~\ref{fig:MSG}). 
This insertion of junctions simplifies recursive scheme for counting HWs: instead of eight different types of walks in the case of GM fractals, one should consider only four types of walks in this case.

Any closed HW within the $(r+1)$th order generator of MSG can be
decomposed into $b(b+1)/2$ strands, each of them being a $B$-type
HW within one of the $b(b+1)/2$ $r$th order triangle inside it.
Here, $B$-type HW has the same meaning as in the case of 3-simplex
lattice, {\em i.e.} it is a HW that enters a generator at one of its
corner vertices, and leaves it through another one, meanwhile
visiting all the remaining vertices, including the third corner
vertex (see Fig.\ ~\ref{fig:MSG}). Recursion relation for the
numbers of $B$-type walks is of the form
\begin{equation}
B^{(r+1)}=q\left(B^{(r)}\right)^{{b(b+1)}\over 2}\, ,
\label{eq:BMSG}
\end{equation}
where $q$ depends only on $b$, and, consequently, for $r>>1$, the number
$B^{(r)}$ behaves as:
\begin{equation}
B^{(r)}\sim \left[q^{\frac{2}{(b-1)(b+2)}}\right]^{\left(
{\frac{b(b+1)}{2}}\right)^{r}}\, .
\end{equation}
The number of closed HWs on the $(r+1)$th order generator
$Z_C^{(r+1)}$ is equal to:
\[
Z_C^{(r+1)}=\mathrm{const}\left(B^{(r)}\right)^{{b(b+1)}\over 2}\, ,
\]
meaning that $Z_C^{(r)}$ has the same asymptotic form as
$B^{(r)}$. Since the number $N_r$ of sites on the $r$th order
generator is equal to $N_r=3[b(b+1)/2]^r$, for large $r$ one gets
the following scaling form:
\begin{equation}
Z_C^{(r)}\sim B^{(r)}\sim \omega^{N_r}\, , \quad \mathrm{with}
\quad \omega=q^{\frac{2}{3(b-1)(b+2)}}  \, . \label{eq:closedMSG}
\end{equation}
Values of $q$ and $\omega$ for $2\leq b\leq 8$ are given in
Table~\ref{table:MSG}.
\begin{table}
\caption{Coefficients $q$, $a$, $c$, $a'$, and $c'$, appearing in
recursion relations (\ref{eq:BMSG}), and (\ref{eq:ACMSG}), for the
numbers of $B$, $A$, and $C$-type HWs, found by direct computer
enumeration of all possible corresponding configurations, together
with the values of connectivity constant $\omega$ and exponent
$\gamma$ for open HWs on MSG fractals with $2\leq b\leq
8$.}\label{table:MSG}
\begin{ruledtabular}
\begin{tabular}{cccccccl}
 $b$ & $q$&$a$ & $c$ & $a'$ & $c'$&$\omega$&$\gamma$ \\\hline
  2 & 1& 2 & 2 & 1&3& 1&2.52372 \\
  3 & 2 &6 & 10 & 4&8&1.04729 &2.12343\\
  4 & 4 & 24 & 42 & 20&40&1.05269 &2.3816 \\
  5 &16 & 134 & 228 &116&196&1.06824 &2.23644 \\
 6 & 68 & 932 &1460  & 886&1408&1.07286& 2.32218 \\
  7 & 464  & 8656 & 12524  & 8372& 11604&1.07875&2.27302  \\
  8& 3838& 101612&133764&103258&133428&1.08177 &2.29907\\
\end{tabular}
\end{ruledtabular}
\end{table}

The number $Z_O^{(r+1)}$ of open HWs on the $(r+1)$th order
generator is equal to:
\[
Z_O^{(r+1)}=B^{\frac{b(b+1)}{2}-2}\left(k_1A^2+k_2AC+k_3C^2+k_4BD\right)\,,
\]
where $A=A^{(r)}$, $B=B^{(r)}$, $C=C^{(r)}$ and $D=D^{(r)}$  have
the same meaning as in the case of 3-simplex lattice, and $k_i$ are
some constants depending only on $b$. Numbers $A$ and $C$ for two
consecutive stages of MSG fractal construction satisfy the recursion
relation:
\begin{equation}\left(
    \begin{array}{c}
      A' \\
      C' \\
    \end{array}
  \right)=\left(B^{(r)}\right)^{\frac{b(b+1)}{2}-1}\left(
                                                     \begin{array}{cc}
                                                       a & c \\
                                                       a' & c' \\
                                                     \end{array}
                                                   \right)\left(
    \begin{array}{c}
      A \\
      C \\
    \end{array}
  \right)\, , \label{eq:ACMSG}
  \end{equation}
where $a$, $c$, $a'$, and $c'$ depend only on $b$, and can be
found by enumeration of the corresponding one-leg HWs (see
Table~\ref{table:MSG}). Dividing this relation by recursion
relation (\ref{eq:BMSG}) for the $B$ numbers, and introducing new
variables: $x_r=A^{(r)}/B^{(r)}$, and $y_r=C^{(r)}/B^{(r)}$, one
gets \begin{equation}\left(
    \begin{array}{c}
      x_{r+1} \\
      y_{r+1} \\
    \end{array}
  \right)={1\over q}\left(\begin{array}{cc}
                                                       a & c \\
                                                       a' & c' \\
                                                     \end{array}
                                                   \right)\left(
    \begin{array}{c}
      x_r \\
      y_r \\
    \end{array}
  \right)\, ,  \label{eq:MSGubaceno}
  \end{equation}
implying that for $r\gg 1$ parameters $x_r$ and $y_r$ behave as
$\lambda^r$, where
\begin{equation}
\lambda={1\over{2q}}\left(a+c'+\sqrt{(a-c')^2+4c\,a'}\right)\,
\label{eq:lambdaMSG}
\end{equation}
is the larger eigenvalue of the 2x2 matrix in
(\ref{eq:MSGubaceno}).

The number $D^{(r)}$ of two-leg HWs transforms as:
\[
D'=B^{\frac{b(b+1)}{2}-2}\left(d_{AA}A^2+d_{AC}\,AC+d_{CC}C^2+d_DBD\right)\, ,\]
which, with $z_r=D^{(r)}/B^{(r)}$ can be rewritten as
\begin{equation}
z_{r+1}={1\over
q}(d_{AA}\,x_r^2+d_{AC}\,x_r\,y_r+d_{CC}\,y_r^2)+{{d_D}\over{q}}\,z_r\,
. \label{eq:zMSG}
\end{equation}
In a similar way as it was done for two-leg walks in the case of
GM fractals (Appendix B) it can be deduced that $d_D/q=b(b+1)/2$.
Then, equation (\ref{eq:zMSG}) implies that the large $r$ behavior
of $z_r$ is governed by the larger of numbers $\lambda^2$ and
$b(b+1)/2$. By computer enumeration of all possible one-leg
configurations we found that inequality $\lambda^2>b(b+1)/2$ is
satisfied for $2\leq b\leq 8$, so that $z_r\sim\lambda^{2r}$.

With the established large $r$ behavior of
$x_r$, $y_r$ and $z_r$, and consequent behavior of $A^{(r)}$, $C^{(r)}$ and $D^{(r)}$,
together with the previously found relation (\ref{eq:closedMSG})
for $B^{(r)}$, it is not difficult to find out that scaling form for
overall number $Z_O^{(r)}$ on MSG fractal is the same as in the case of
corresponding GM fractal (\ref{eq:openGM}), but with different
values of $\omega$ and $\lambda$. Consequently, the only exception - equality of the values of $\gamma$ for 3-simplex and SG fractals ($b=2$ case of MSG and GM fractals, respectively), seems to be accidental. Indeed, this exception is also an indication that number of contacts per monomer is not the only relevant physical measure of HWs frustration.

\begin{figure}
\includegraphics[height=80mm]{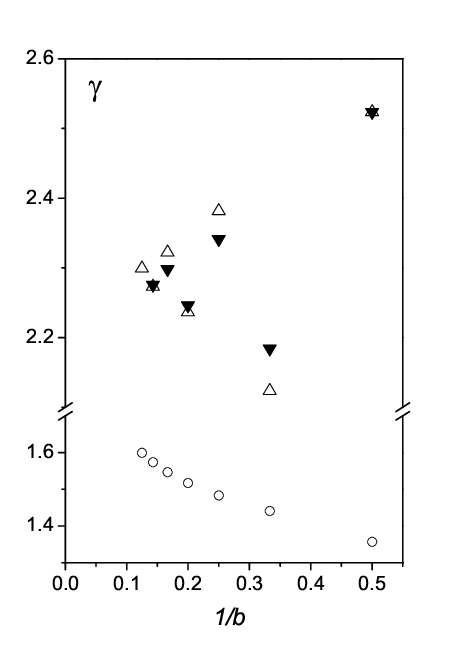}
\caption{Values of the exponent $\gamma$ appearing in the scaling
form $Z_O\sim\omega^{N}N^{\gamma}$ found for the number of open
HWs on GM (full triangles) and MSG ($\triangle$) fractals,
together with the corresponding values for SAWs (o) on the same
fractal families \cite{EKM}, as functions of the reciprocal of the
fractal parameter $b$.}
 \label{fig:GrafikGMMSG}
\end{figure}

The particular values of the connectivity constant $\omega$ and the exponent $\gamma$ for
MSG fractals with $2\leq b\leq 8$ are given in Table~\ref{table:MSG}, whereas in Fig.\ ~\ref{fig:GrafikGMMSG} we depict $\gamma$ as a function of $1/b$, for both GM and MSG fractals. One can see that $\gamma$ is not a monotonic function of $b$, which has not been found for any of the critical exponents connected with SAWs on these fractal families \cite{EKM,ZivicGamaSG,ZivicNiSG}. For the sake of comparison on the same figure we give the corresponding values of $\gamma$ for open SAWs \cite{EKM}, which are the same for GM and MSG fractals with equal $b$. In addition, one can also observe that values of $\gamma$ for HWs are separated in two groups: for even values of $b$ exponent $\gamma$ decreases when $b$ grows, whereas for $b$ odd $\gamma$ increases with $b$. This, however, resembles the behavior of the exponent $\psi$, appearing in the scaling form ${\lambda^N}\mu^{N^{\psi}}$ for the average number of different configurations of a branched polymer of $N$ bonds on GM fractal with $b>2$. Namely, for even and odd values of $b$ in a recent paper \cite{DharBranched} Dhar obtained two different exact expressions for $\psi$. Finally, it is clearly seen in Fig.\ ~\ref{fig:GrafikGMMSG} that the difference between $\gamma$ values for GM and MSG fractals becomes smaller with larger $b$, so it seems plausible to investigate HWs only on MSG fractals (which is simpler than GM case), in order to obtain the large $b$ behavior of the exponent $\gamma$. This limit is interesting because for large $b$ already at the first stage of the construction of the GM fractal, one gets finite, but large, homogeneous triangular lattice, for which asymptotic form of the number of HWs has not been established yet.

\section{n-simplex lattices}\label{sec:nsimplex}

In order to extend our analysis to higher dimensional lattices, in
this section we turn to $n$-simplex fractals with $n>3$, which are
embedded in $d=n-1$ dimensional Euclidean spaces. To obtain an
$n$-simplex lattice \cite{DharNSimplex} one starts with a complete
graph of $n$ points and replaces each of these points by a new
complete graph of $n$ points. The subsequent stages are
constructed self-similarly, by repeating this procedure. After $l$
such iterations one obtains an $n$-simplex of order $l$, which
consists of $N_l=n^l$ points.   The complete $n$-simplex lattice
is obtained in the limit $l\to\infty$. Fractal dimension $d_f$ of
$n$-simplex lattice is equal to $d_f=\ln n/\ln 2$.

\subsection*{Closed HWs}

\begin{figure}
\includegraphics[width=80mm]{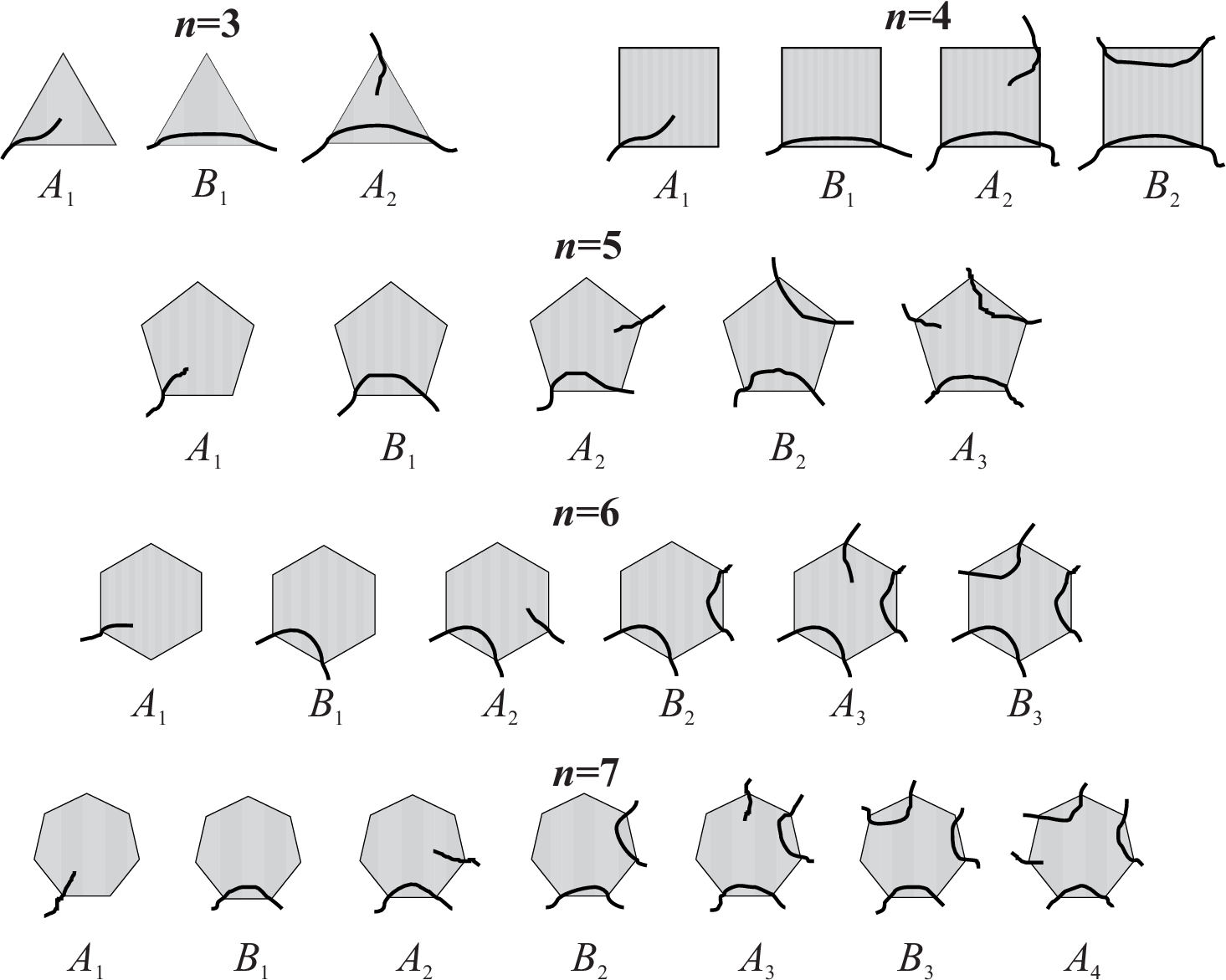}
\caption{All types of HWs needed for determining the connectivity
constants and asymptotic forms of the numbers of HWs for
$n$-simplex lattices with $3\leq n\leq 7$. Gray polygons denote
$n$-simplex of order $l$.}
 \label{fig:nsimplexParametri}
\end{figure}

Any closed HW on $n$-simplex of order $l+1$ can be decomposed into
$n$ parts within its $n$ simplices of order $l$. Parts (steps) of
the walk within the simplices of order $l$ can be of $[n/2]$
different types (see Fig.\ ~\ref{fig:nsimplexParametri}),
and we shall denote the numbers of corresponding HWs by $B_1$,
$B_2$, $\ldots$, $B_{[n/2]}$. The overall number of closed HWs is equal
to
\begin{equation}
Z_C^{(l+1)}=\sum\limits_{i_1+\cdots+ i_{[\frac{n-1}2]}=n}\, c_{i_1\,
i_2\ldots i_{[\frac{n-1}2]}}\prod\limits_{j=1}^{[\frac{n-1}2]}\,
B_j^{i_j}\, ,\label{eq:nSimplexZatvorene}
\end{equation}
where $c_{i_1\, i_2\cdots i_{[(n-1)/2]}}$ is the number of closed HW
configurations with $i_1$, ..., $i_{[(n-1)/2]}$ steps of $B_1$, ...,
$B_{[(n-1)/2]}$ type, respectively. Numbers $B_i$ satisfy a closed set
of recursion relations:
\begin{eqnarray}
B_i^{(l+1)}&=&\sum\limits_{i_1+\cdots+ i_{[n/2]}=n}\,
b^{(i)}_{i_1\, i_2\cdots i_{[n/2]}}\prod\limits_{j=1}^{[n/2]}\,
B_j^{i_j}\, ,\nonumber\\* i&=&1,2, \cdots, [n/2] \, ,
\label{eq:nSimplexTraversing}
\end{eqnarray}
where $b^{(i)}_{i_1\, i_2\cdots i_{[n/2]}}$ is the number of HW
configurations of $B_i$-type that traverse the simplex of order
$(l+1)$, and have $i_1$, ..., $i_{[n/2]}$ steps of $B_1$, ...,
$B_{[n/2]}$ type, respectively.  Due to the self-similarity of the
underlying lattice both $c_{i_1\, i_2\cdots i_{[(n-1)/2]}}$ and
$b^{(i)}_{i_1\, i_2\cdots i_{[n/2]}}$ depend only on $n$. Starting
with the initial values $B_1^{(0)}=1$,
$B_2^{(0)}=\cdots=B_{[n/2]}^{(0)}=0$, one can iterate recursion
relations (\ref{eq:nSimplexTraversing}) and calculate $B_i^{(l)}$
and $Z_C^{(l)}$, in principle for any $l$, but since all these
numbers grow very fast with $l$, it is useful to introduce new
parameters: $x_i=B_i/B_{[n/2]}$, $i=1$, ..., $[n/2]-1$. Dividing
relations for $B_i^{(l+1)}$ by $B^{(l+1)}_{[n/2]}$ one obtains new
recursion relations:
\begin{equation}
x_i^{(l+1)}=\frac{f_i\left(x_1^{(l)},\ldots,x^{(l)}_{[n/2]-1}\right)}{f\left(x_1^{(l)},
\ldots\,x^{(l)}_{[n/2]-1}\right)}\, ,
\label{eq:nSimplexX}\end{equation} where
\begin{eqnarray}
&&f_i(x_1^{(l)},\ldots,x_{[n/2]-1}^{(l)})=B_i^{(l+1)}/\left(B_{[n/2]}^{(l)}\right)^n,
 \nonumber\\
&&f(x_1^{(l)},\ldots\,x_{[n/2]-1}^{(l)})=B_{[n/2]}^{(l+1)}/\left(B_{[n/2]}^{(l)}\right)^n, \label{eq:funkcijaf}
\end{eqnarray}
$i=1,\ldots,[n/2]-1$. Iterating these relations (starting with
$l=1$) one notices that large $l$ behavior of $x_i^{(l)}$ strongly
depends on the parity of $n$. It turns out that for odd $n$ values all $x_i^{(l)}$
tend to some finite constants, whereas for even $n$ the following
asymptotic relations are valid:
\begin{equation} x_i^{(l)}\sim\,
\left({\lambda_B}^{k-i}\right)^{2^l}\, , \quad i=1,\cdots, k-1\, ,
\label{eq:lambdab}
\end{equation}
where $k=[n/2]$ and $\lambda_B$ is some finite constant less than
1. Such behavior is a direct consequence of the fact that for even values of $n$ none of $B_1$, $B_2$, $\ldots$, $B_{k-1}$-type configurations within a simplex of order $(l+1)$ can be accomplished by $n$ steps of $B_{k}$-type through comprising $n$ simplices of order $l$ (see Appendix~\ref{ap:nsimplex}), which is not the case for simplices with odd $n$, because $B_i$-type steps occupy even number of corner vertices of any simplex.

With the established large $l$ behavior of numbers $x_i$ it can be shown that from (\ref{eq:nSimplexZatvorene}) follows
\begin{equation}
Z_C^{(l+1)}\, \sim \cases{ \left(B_k^{(l)}\right)^n \, , &
for $n$ odd, \cr \left(\lambda_B^{2^l}B_k^{(l)}\right)^n \, ,& for
$n$ even.\label{eq:ZatvoreneN}}
\end{equation}
Although these two asymptotic forms are not the same, from the
definition (\ref{eq:definicija}) of the connectivity constant
$\omega$ in both cases one gets
\[
\ln\omega=\lim\limits_{l\to\infty}\frac{\ln
Z_C^{(l+1)}}{n^{l+1}}=\lim\limits_{l\to\infty}\frac{\ln
B_{[n/2]}^{(l)}}{n^{l}}\, .
\]
The last limiting value can be obtained via numerical iteration of
the recursion relation
\begin{equation}
\frac{\ln B_{k}^{(l+1)}}{n^{l+1}}=\frac{\ln
B_{k}^{(l)}}{n^{l}}+{1\over{n^{l+1}}}\ln
f(x_1^{(l)},\ldots\,x_{k-1}^{(l)})\, , \label{eq:IteracijaZaOmega}
\end{equation}
which follows directly from (\ref{eq:funkcijaf}). Furthermore, from (\ref{eq:IteracijaZaOmega}) one has
\[
\frac{\ln B_k^{(p)}}{n^{p}}=\frac{\ln
B_k^{(l)}}{n^{l}}+\sum\limits_{m=l}^{p-1}{1\over{n^{m+1}}}\ln
f(x_1^{(m)},\ldots ,x_{k-1}^{(m)}),
\]
which,  with
$p\to\infty$, gives \[
\ln\omega=\frac{\ln
B_k^{(l)}}{n^{l}}+\sum\limits_{m=l}^{\infty}{1\over{n^{m+1}}}\ln
f(x_1^{(m)},\ldots\,x_{k-1}^{(m)})\, .
\]
Then, since $f(x_1^{(m)},\ldots\,x_{k-1}^{(m)})$ decreases with
$m$, it follows that $n^l\ln\omega -\mathrm{const}<\ln
B_k^{(l)}<n^l\ln\omega$, {\em i.e.} $\ln B_k^{(l)}\sim n^l\ln\omega$.
Using this, together with (\ref{eq:ZatvoreneN}), for the numbers of closed HWs  one finally obtains the scaling formulas
\begin{equation}
Z_C^{(l)}\sim \cases{
    \omega^{N_l}\, , & for $n$ odd, \cr
    \omega^{N_l}\,\left(\lambda_B^{n/2}\right)^{N_l^\sigma}\, , & for $n$ even,}\label{eq:scalingClosed}
\end{equation}
 where $\sigma=\frac{\ln 2}{\ln
n}={1\over{d_f}}$, with $d_f$ being the fractal dimension of the
$n$-simplex lattice. Scaling forms and particular values for
$\omega$ were found in \cite{Bradley} for $n=4$, in \cite{Stajic}
for $n=5$, and $6$, and for $n=7$ in Appendix~\ref{ap:nsimplex} of this paper.

\subsection*{Open HWs}

Any open HW on $n$-simplex of order $(l+1)$ can be decomposed into
$n$ parts within its $n$ simplices of order $l$. The two ends of
the HW are either both in the same or in two different $l$th order
$n$-simplices. In the first case the configuration of the
corresponding HW part is of the two-leg type, whereas in the
latter case the configurations of both ending parts are of
the one-leg type. For each $n$ there are $n-[n/2]$ different types
of one-leg configurations (for $3\leq n\leq 7$ they are depicted
in Fig.\ ~\ref{fig:nsimplexParametri}). The remaining $(n-1)$ or
$(n-2)$ parts of the open HW correspond to $B_i$-type steps. If we denote the numbers of
one- and two-leg walks by $A_i^{(l)}$ and $D_i^{(l)}$,
respectively, then the overall number of open HWs on $n$-simplex
of order $(l+1)$ is equal to
\begin{equation}
Z_O^{(l+1)}=\sum\limits_{i,j=1}^{n-\left[{n\over 2}\right]}
E_{ij}A^{(l)}_iA^{(l)}_j+\sum\limits_{i=1}^{\left[{n\over
2}\right]}F_iD^{(l)}_i\, , \label{nSimplexOtvorene}
\end{equation}
where
\begin{eqnarray}
E_{ij}&=&\sum\limits_{\sum\limits_{r=1}^{[n/2]}i_r=n-2}\,
e^{(ij)}_{i_1\, i_2\cdots
i_{\left[{n/2}\right]}}\prod\limits_{r=1}^{\left[{n/2}\right]}\, \left(B_r^{(l)}\right)^{i_r}, \nonumber\\
F_i&=&\sum\limits_{\sum\limits_{r=1}^{[n/2]}i_r=n-1}\,
f^{(i)}_{i_1\, i_2\cdots
i_{\left[{n/2}\right]}}\prod\limits_{r=1}^{\left[{n/2}\right]}\,
\left(B_r^{(l)}\right)^{i_r},
\end{eqnarray}
with coefficients $e$ and $f$ being the numbers of corresponding
HW configurations. Numbers $A_i^{(l)}$ of one-leg walks
satisfy recursion relations which can be put in the following
matrix form
\begin{equation}
\left(
  \begin{array}{c}
    A_1^{(l+1)} \\
    \vdots \\
    A^{(l+1)}_{p} \\
  \end{array}
\right)=\left(
          \begin{array}{ccc}
            a_{11} &  \cdots & a_{1p} \\
                       \vdots &  \ddots & \vdots \\
            a_{p1} &
            \cdots & a_{pp} \\
          \end{array}
        \right)\left(
                 \begin{array}{c}
                   A_1^{(l)} \\
                   \vdots \\
                   A_{p}^{(l)} \\
                 \end{array}
               \right)\, , \quad  \label{eq:Jednokrake}
\end{equation}
where $p=n-[n/2]$, and
\begin{equation}
a_{ij}=\sum\limits_{\sum\limits_{r=1}^{[n/2]}i_r=n-1}\,
\alpha^{(ij)}_{i_1\, i_2\cdots
i_{\left[{n/2}\right]}}\prod\limits_{r=1}^{\left[{n/ 2}\right]}\,
\left(B_r^{(l)}\right)^{i_r}\, ,\label{eq:alfa}
\end{equation}
with $\alpha^{(ij)}_{i_1\, i_2\cdots i_{\left[{n/2}\right]}}$
being constant coefficients. Introducing parameters
$y_i=A_i/B_{[n/2]}$, $i=1, ..., p$, and dividing relations for
$A_i^{(l+1)}$ by $B^{(l+1)}_{[n/2]}$ one gets recursion relations
for the new parameters:
\begin{equation}
\left(
  \begin{array}{c}
    y_1^{(l+1)} \\
       \vdots \\
    y_{p}^{(l+1)} \\
  \end{array}\right)=\left(
                \begin{array}{cccc}
                  m_{11} &  \cdots & m_{1p} \\
                  \vdots &  \ddots & \vdots \\
                  m_{p1} & \cdots &
                  m_{pp} \\
                \end{array}
              \right)\left(
                       \begin{array}{c}
                         y_1^{(l)} \\
                                                 \vdots \\
                         y_{p}^{(l)} \\
                       \end{array}
                     \right)\, ,  \label{eq:YJednokrake}
\end{equation}
where \[
m_{ij}=\frac{a_{ij}(B_1,\ldots,B_{[n/2]})}{f(x_1,\ldots,x_{k-1})(B_{[n/2]})^{n-1}}=
\frac{g_{ij}(x_1,\ldots,x_{k-1})}{f(x_1,\ldots,x_{k-1})}\, ,
\]
with $g_{ij}$ being polynomials of order
$n-1$, whose coefficients depend only on $n$, and $f$ defined
in (\ref{eq:funkcijaf}). Numerically iterating relations
(\ref{eq:nSimplexX}) for $x_i$, and substituting them into
functions $m_{ij}$, after large number of iterations one obtains
constant values, such that \begin{itemize} \item for odd $n$,
matrix $m_{ij}$ has only one eigenvalue $\lambda$ larger than 1,
implying that $y_i^{(l)}\sim \lambda^l$, \item  for even values of
$n$, only $m_{pp}\to 1$, whereas all other $m_{ij}\to 0$,
implying that $y_{p}^{(l)}$ tends to finite constant value, whereas
all the other $y_i^{(l)}$ tend to 0.
\end{itemize}
Knowing the behavior of parameters $y_i^{(l)}$, one can get the
large $l$ behavior of the numbers $A_i^{(l)}$, and, consequently,
the asymptotic form of the first sum in the right-hand side of
(\ref{nSimplexOtvorene}). In addition, it can be shown that
numbers of two-leg configurations $D_i$ are not necessary for establishing the
asymptotical behavior of (\ref{nSimplexOtvorene}) (terms with $D_i^{(l)}$ either scale as terms with $A_i^{(l)}A_j^{(l)}$ or they are much smaller).
For $n$-simplex lattices with odd $n$ all terms with $A_i^{(l)}A_j^{(l)}$ have the same asymptotical behavior, whereas for even $n$ behavior of $Z_O^{(l+1)}$ is governed by the term with $(A_p^{(l)})^2$. In both cases the following asymptotical formula is valid:
\[
Z_O^{(l+1)}\sim \left(y_p^{(l)}\right)^2 \left(x_{[n/2]-1}^{(l)}\right)^{n-2}\left(B_{[n/2]}^{(l)}\right)^n\, ,
\]
which, with obtained behavior of $y_p^{(l)}$ and $x_{[n/2]-1}^{(l)}$, directly gives:
\begin{equation}
Z_O^{(l)}\sim\cases{\omega^{N_l}\, N_l^\gamma\, , \quad
\gamma=2\frac{\ln\lambda}{\ln n}\, ,& for odd $n$,\cr
\omega^{N_l}\,\left(\lambda_B^{\frac{n-2}{2}}\right)^{N_l^\sigma}\!\!\!, \,\sigma=\frac{\ln 2}{\ln n}, & for even $n$.}\label{eq:scalingOpen}
\end{equation}
Details of the derivation of these formulas can be found in
Appendix~\ref{ap:nsimplex}. Value of $\gamma=2.5237...$ for $n=3$ was found in Sec.~II,
whereas for $n=5$ and 7, exponent $\gamma$ is equal to
2.1668... and 2.1079..., respectively.

 As in GM and MSG fractal cases, one observes that behavior of the number of long HWs on $n$-simplex lattices  is strongly affected by the parity of the fractal parameter. For lattices with odd $n$ scaling forms are the same as for GM and MSG fractals. This scaling form coincides with the scaling forms for average numbers of $N$-bonded SAWs on these lattices: with commonly accepted symbols for critical exponents, average number of closed SAWs scales as $\mu^NN^{\alpha-3}$, whereas for the average number of open SAWs formula $\mu^NN^{\gamma-1}$ is valid (these are the same formulas as for SAWs on homogeneous lattices). Comparing these formulas with those obtained for HWs (and keeping in mind that formulas for HWs correspond to the overall numbers of HWs), one can say that corresponding value $\alpha=2$, obtained for closed HWs on all these lattices, certainly differs from the values obtained for SAWs \cite{EKM,DharNSimplex,KSJ}. The same holds for exponent $\gamma$, meaning that HWs and SAWs belong to different universality classes (see Fig.\ ~\ref{fig:GrafikGMMSG}). For HWs on $n$-simplex lattices with even $n$ we obtained different scaling forms:  $\omega^N(\lambda_B^{n/2})^{N^\sigma}$ for closed, and $\omega^N(\lambda_B^{(n-2)/2)})^{N^\sigma}$ for open walks. However, one should note here that, whereas polymers on GM, MSG, and $5$-simplex fractals for $T>0$ can exist only in swollen phase, on 4- \cite{DharVannimenus}, and 6-simplex \cite{KumarSingh} lattices,  bellow some finite temperature $T_\theta>0$ polymers collapse into a more compact phase. Scaling of collapsed polymers has not been analyzed so far, and behavior of HWs might be an indication that on $n$-simplex lattices with even $n$, for $T<T_\theta$ polymers do not scale in the same manner as they do for $T>T_\theta$.

\section{Summary and conclusion}

In this paper we have analyzed asymptotic behavior of the numbers of open and closed  Hamiltonian walks on Given-Mandelbrot, modified Sierpinski gasket and $n$-simplex fractal families. Obtained scaling forms are summarized in Table~\ref{table:ScalingForms}.
\begin{table}
\caption{Asymptotic behavior of the numbers of open and closed numbers of HWs on lattices with $N$ sites. Values of $\omega$, $\lambda_B$, and $\gamma$ are given in the text, and $d_f$ is the fractal dimension of the corresponding lattice.}\label{table:ScalingForms}
\begin{ruledtabular}
\begin{tabular}{ccc}
 Lattice& Closed HWs&Open HWs  \\\hline\hline
  all GM and MSG fractals \\ $n$-simplex with odd $n$&$\omega^N$&$\omega^N N^\gamma$ \\\hline\\
  $n$-simplex with even $n$& $\omega^N\left(\lambda_B^{\frac n2}\right)^{N^{ 1/{d_f}}}$&$\omega^N \left(\lambda_B^{\frac{n-2}2}\right)^{N^{1/{d_f}}}$\\
  \end{tabular}
\end{ruledtabular}
\end{table}
One can see that scaling form (\ref{eq:asimptotika}): $Z_N\sim \omega^N {\mu^{N^\sigma}_S}N^a$, proposed and obtained for some homogeneous lattices with large number of sites $N$, is satisfied for all fractal lattices under consideration, but,
with at least one of the exponents $\sigma$ and $a$ being equal to 0. For HWs on GM, MSG, and $n$-simplex lattices with odd $n$ scaling form is $\omega^N$ for closed, and  $\omega^N N^\gamma$ ($\gamma>0$) for open walks, {\em i.e.} exponent $\sigma$ formally may be taken as 0. This value cannot be obtained with a simple generalization of the formula $\sigma=(d-1)/d$, valid for $d$-dimensional homogeneous lattice, but it can be explained following the original argument of Owczarek {\em et al} \cite{Owczarek}, which led to (\ref{eq:asimptotika}) for collapsed polymers on homogeneous lattices. Namely, collapsed polymer chain on a homogeneous lattice has a form of a compact globule, with a sharp boundary separating it from the surrounding solvent. Monomers on the boundary have smaller number of contacts with other monomers then those in the bulk of the globule, so that a factor $\mu_S^{N_S}$, with $N_S$ being the number of monomers on the boundary, should appear in the scaling form. Furthermore, if one assumes that the boundary itself is homogeneous surface, then $N_S\sim N^{(d-1)/d}$, where $N$ is the number of monomers in the polymer chain.  The crucial point of this argument is existence of two sets of monomers with different number of contacts. For HWs on both 3-simplex lattice and Sierpinski gasket (MSG and GM with $b=2$, respectively), monomers at all sites, except for the three outer vertices of the lattices, have the same number of contacts (see Figs~ \ref{fig:3simplexHW} and \ref{fig:SGb2HW}), so that 'surface' factor with nontrivial $\sigma$ cannot arise.  For GM fractals with $b>2$ it can be shown that on the $r$th stage of fractal construction there are $N_B^{(r)}=\frac{b-2}b\{[b(b+1)/2]^r-1\}$ 'bulk' monomers with four contacts, whereas $N_S^{(r)}-3=N^{(r)}-N_B^{(r)}-3$ 'surface' monomers, with $N^{(r)}$ given by (\ref{eq:GMcvorovi}), have two contacts. This means that for $r\gg 1$ both $N_B^{(r)}$ and $N_S^{(r)}$ scale as $N^{(r)}\sim [b(b+1)/2]^r$, so that $\mu_S^{N_S}$ can be incorporated into the term $\omega^N$, or formally one can put $\sigma=0$.  For HWs on MSG fractals with $b>2$, as well as on $n$-simplex lattices with odd $n$,  all sites have the same coordination number (except for the finite number of corner vertices of the whole lattice), and consequently all monomers have the same number of contacts. Hence, the argument of Owczarek {\em et al} for these lattices also gives $\sigma=0$. However, presence of the stretched-exponential term $(\lambda_B^{n/2})^{N^{1/d_f}}$ for closed, and $(\lambda_B^{(n-2)/2)})^{N^{1/d_f}}$ for open Hamiltonian walks  on $n$-simplex lattices with even $n$, can't be a consequence of surface effects, since all the sites of these lattices also have the same coordination number. This means that there is some additional effect that should be considered in order to explain obtained  scaling forms. It is not clear what that effect is, but one should notice that for $l$th order $n$-simplex $N_l^{1/d_f}=2^l$, which is equal to the linear size of the lattice. Also, some clue about it can be achieved by careful inspection of the differences between the asymptotic behavior of the requisite numbers of HWs for lattices with even and odd values of $n$. For instance, for odd $n$ the numbers of long $B_i$-type HWs fulfil relation:  $B_1^{(l)}\sim B_2^{(l)}\sim\cdots\sim B^{(l)}_{[n/2]}$, whereas for even $n$ the number $B^{(l)}_{n/2}$ is much larger than any other $B^{(l)}_i$. Hence, one can expect stretched-exponential terms in scaling forms for HWs on lattices on which 'entangled' configurations are more probable. This bears some analogy with square lattice, for which  it was found that large fraction of monomers participate in secondary structures \cite{proteinfolding}.

To conclude, we can say that exact recursive technique, used in this paper for enumeration of all possible Hamiltonian walks on Given-Mandelbrot, modified Sierpinski gasket and $n$-simplex fractal lattices, proved to be very efficient for open, as well as for closed walks. Recursion relations for the number of different types of compact configurations in simpler cases could be obtained by direct enumeration of the corresponding walks, whereas for lattices with more complicated structure we used appropriate computer programs. Once recursion relations were established, it was possible to generate and analyze numbers of very long HWs, eventually obtaining their asymptotical behavior. We believe that obtained scaling forms can help in better understanding of real compact polymers, therefore we intend to apply this technique to HWs on fractals embedded in three-dimensional space, in particular on 3d generalizations of GM and MSG fractals. It would also be interesting to extend the method to interacting HWs. Finally, we hope that our results might be useful in finding scaling forms for collapsed phase of interacting self-avoiding walks on fractals for which transition from swollen to collapsed phase occurs at $T_\theta>0$.

\begin{acknowledgments}
We would like to thank I. \v Zivi\' c for careful and critical reading of the manuscript.  SEH and SM acknowledge the financial support
from the Serbian Ministry of Science and Environmental Protection
(Projects No: OI 141020B and OI 144022). \end{acknowledgments}

\appendix

\section{Recursion relations for one-leg HW{\lowercase{s}} on GM fractals
\label{dod:JednokrakeGM}}

\begin{figure*}
\includegraphics[width=130mm]{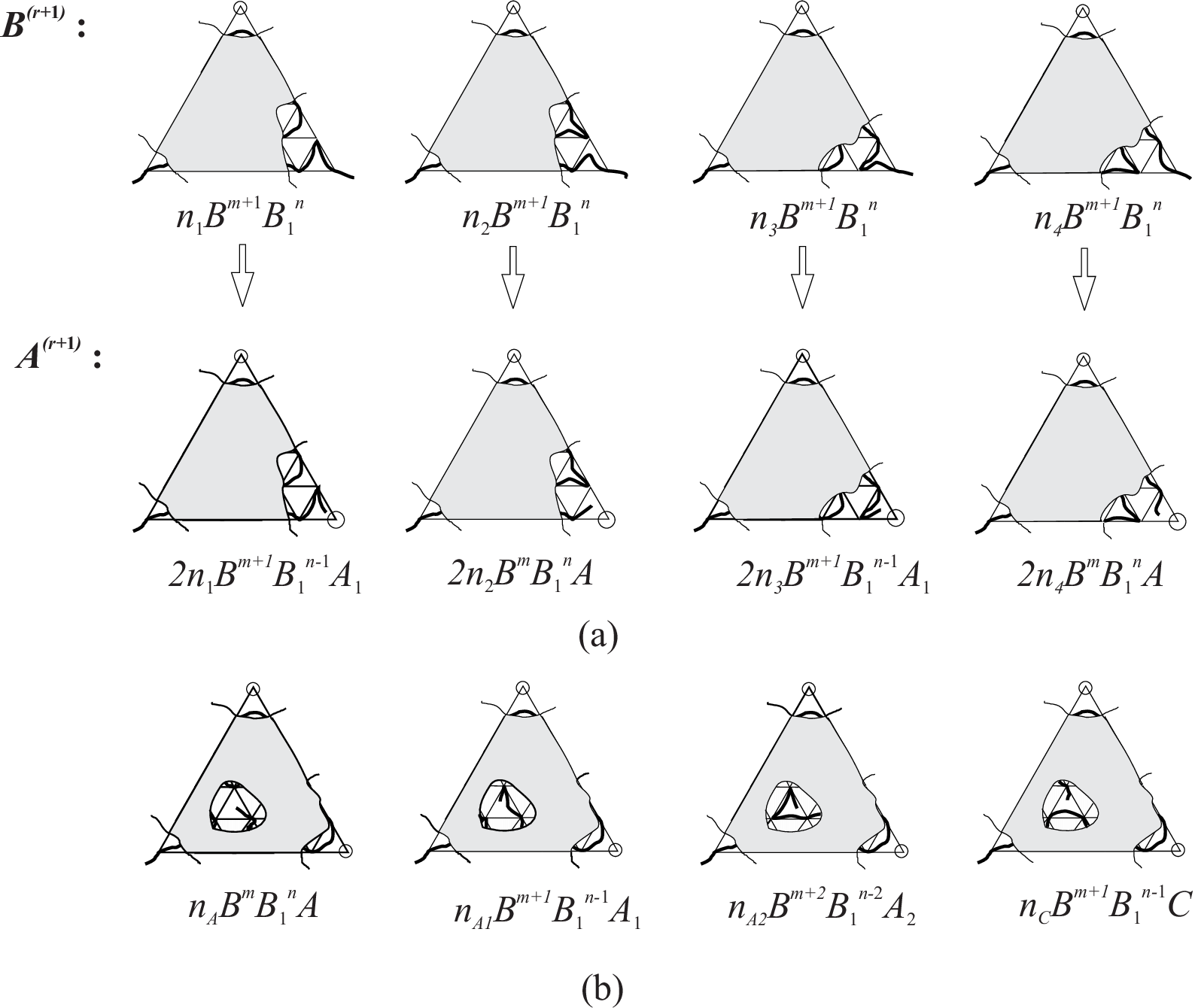}
\caption{(a) Possible configurations of $B$-type HWs on $(r+1)$th
order generator of GM fractal,  and corresponding $A$-type HWs,
together with the corresponding numbers of walks. Small
upward-oriented triangles represent the $r$th order generators.
Encircled nodes are the corner vertices which are not visited by
the corresponding HW. (b) Possible types of $A$-type HWs which
terminate in interior $r$th order
generators.\label{fig:AppA1}}\end{figure*}

In this Appendix we prove that numbers of one-leg HWs:
$A$, $A_1$, $A_2$, and $C$, on Given-Mandelbrot fractals, satisfy
recursion relations of the form: \begin{widetext}
\begin{eqnarray}
A'&=&aAB^{m}B_1^{n}+a_1A_1B^{m+1}B_1^{n-1}+a_2A_2B^{m+2}B_1^{n-2}+
\,cB^{m+1}B_1^{n-1}C\, ,\nonumber\\
A_1'&=&\left(a-\frac{p}{2}\right)AB^{m-1}B_1^{n+1}+
a_1A_1B^{m}B_1^{n}+(a_2+p)A_2B^{m+1}B_1^{n-1}+\,cB^{m}B_1^{n}C\, ,\label{eq:onelegGM}\\
A_2'&=&\frac{a-p}{2}AB^{m-2}B_1^{n+2}+\frac{a_1}{2}A_1B^{m-1}B_1^{n+1}+
\left(\frac{a_2}{2}+p\right)A_2B^{m}B_1^{n-1}+
\frac{c}{2}B^{m-1}B_1^{n+1}C\, ,\nonumber\end{eqnarray}\end{widetext}
where $p$ is the number of $B$-type (or $B_1$-type) configurations
within the $(r+1)$th order generator (see (\ref{eq:bezkrake})),
whereas $a$, $a_1$, $a_2$, and $c$ are some integers depending
only on the scaling parameter $b$ of the fractal, and
$m=b(b-1)/2$, $n=b-1$. In order to do that, we first notice that
all configurations of $B$-type can be divided in four sets,
regarding the last step (``step'' is here a part of the walk within
the $r$th-order generator), as is sketched in
Fig.\ ~{\ref{fig:AppA1}}(a). The number of
configurations within each set is denoted by $n_i$, so that
$n_1+n_2+n_3+n_4=p$. Each of the configurations from any of these
four sets can be transformed into one $A$-type configuration by
cutting the last step. In such a way, if the last step was
$B$-type it is converted into an $A$-type step, whereas a
$B_1$-type step is converted into an $A_1$-type step. Number of
configurations obtained in this manner, contributing to
$A^{(r+1)}$, should be doubled, due to symmetry (there are two
corner $r$th order generators in which $A$-type walk can
terminate). This means that
$(n_2+n_4)(B^{(r)})^{m+1}(B_1^{(r)})^{n}$ HWs of $B$-type on
$(r+1)$th order generator can be transformed into
$2(n_2+n_4)(B^{(r)})^{m}(B_1^{(r)})^{n}A^{(r)}$ HWs of $A$-type on
the same generator. The remaining
$(n_1+n_3)(B^{(r)})^{m+1}(B_1^{(r)})^{n}$ HWs of $B$-type on the
$(r+1)$th order generator can be converted into
$2(n_1+n_3)(B^{(r)})^{m+1}(B_1^{(r)})^{n-1}A_1^{(r)}$ HWs of
$A$-type (Fig.\ ~{\ref{fig:AppA1}}(a)).

In addition, an $A$-type configuration can have its last step in
an interior $r$th order generator, and, as can be seen in Fig.\ ~{\ref{fig:AppA1}}(b), that step can be of any of the four possible one-leg
types. Number of such walks with the $A$-type end is of the form
$n_A(B^{(r)})^{k}(B_1^{(r)})^{l}A^{(r)}$, where $k$ and $l$ are
numbers of $B$ and $B_1$-type steps, respectively. Numbers $k$ and
$l$ must satisfy equation $k+l+1=b(b+1)/2-2$, because every $r$th
order triangle must be traversed by HW, and also
$k+2l+2=(b+1)(b+2)/2-2$, because every corner vertex of every
$r$th order generator within the $(r+1)$th order generator, with
the exception of two outer vertices, must be visited. The only
solution of these two equations is $k=m$, $l=n$. In a similar way
one can obtain the numbers of walks with the $A_1$, $A_2$ or
$C$-type end in interior $r$th order generators, so that the
number $A^{(r+1)}$ is equal to
\begin{eqnarray}
A^{(r+1)}&=&(2n_1+2n_3+n_{A1})B^{m+1}B_1^{n-1}A_1\nonumber\\
&+&(2n_2+2n_4+n_A)B^{m}B_1^nA\nonumber\\*&+&
n_{A2}B^{m+2}B_1^{n-2}A_2+n_{C}B^{m+1}
B_1^{n-1}C\,.\nonumber\end{eqnarray} On the other hand, from
Fig.\ ~\ref{fig:AppA1}(a) it is quite obvious  that $n_1=n_2$ and
$n_3=n_4$, implying that $n_1+n_3=n_2+n_4=p/2$, and consequently:
\begin{eqnarray}
A^{(r+1)}&=&(p+n_A)B^{m}B_1^nA+
(p+n_{A1})B^{m+1}B_1^{n-1}A_1\nonumber\\
&+&n_{A2}B^{m+2}B_1^{n-2}A_2+n_{C}B^{m+1}B_1^{n-1}C\,.\label{eq:recA}
\end{eqnarray}

In a similar way, both $A_1$ and $A_2$-type configurations within
the $(r+1)$th order generator can be divided in two classes: (1)
configurations terminating in some interior $r$th order generator, and
(2) configurations terminating in one of the two possible corner
$r$th order generators. The number of $A_1$-type configurations in
the first class is the same as in the corresponding case of
$A$-type configurations, and the only difference between such $A$
and $A_1$-type configurations is that $B$-step through one corner
triangle should be replaced by $B_1$-step. These configurations
then give rise to terms
$n_AB^{m-1}B_1^{n+1}A+n_{A1}B^{m}B_1^{n}A_1
+n_{A2}B^{m+1}B_1^{n-1}A_2+n_{C}B^{m} B_1^{n}C$ in the relation
for $A_1^{(r+1)}$. The number of $A_2$-type configurations in the
first class can be obtained from the corresponding $A$-type
configurations by dividing their number by two (because it is
predefined in which order the corner vertices are visited) and
substituting the two corner $B$-steps by $B_1$ steps, thus leading
to terms $(n_AB^{m-2}B_1^{n+2}A$ $+n_{A1}B^{m-1}B_1^{n+1}A_1$
$+n_{A2}B^{m}B_1^{n-1}A_2$ $+n_{C}B^{m-1} B_1^{n+1}C)/2$ in the
relation for $A_2^{(r+1)}$.

\begin{figure*}
\includegraphics[width=130mm]{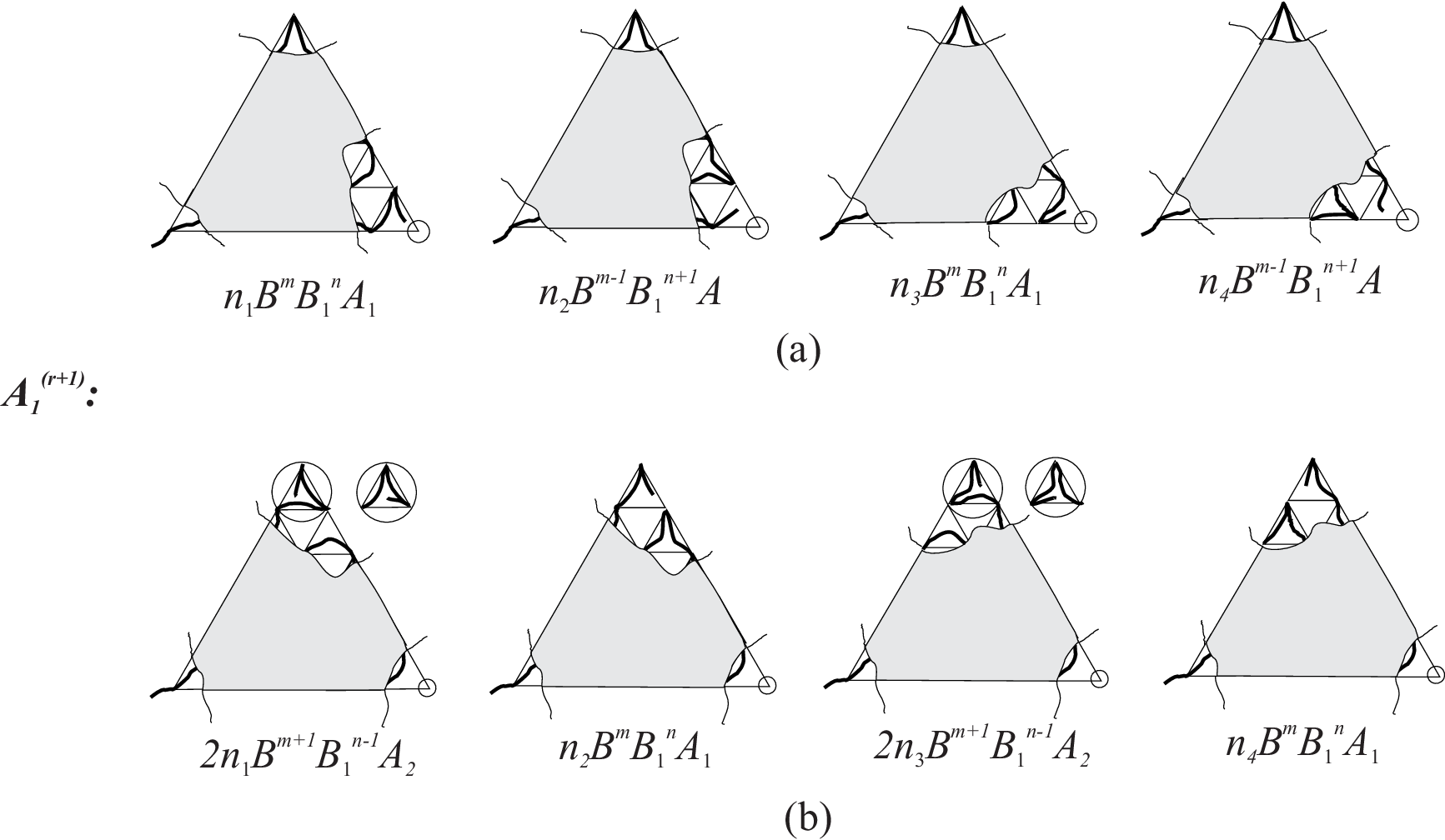}
\caption{Possible configurations of $A_1$-type HWs on $(r+1)$th
order generator of GM fractal, which terminate in (a) the
right-corner and (b) the upper corner $r$th order generator,
together with the corresponding numbers of walks.
\label{fig:AppA2}}\end{figure*}

As for the second class, lets first consider $A_1$-type
configurations which visit the upper corner vertex
(Fig.\ ~\ref{fig:AppA2}(a)), and terminate in the right corner
triangle. Each of them can be obtained from exactly one $B_1$-type
configuration, by converting the last $B$ or $B_1$-step through
the right corner triangle into $A$ or $A_1$-step, in a similar way
as was done in the case of $A$-type configurations ($B_1$-type
configurations can be partitioned in the same way as $B$-type
configurations in Fig.\ ~\ref{fig:AppA1}(a), with the only
difference that $B$-step through the upper corner triangle is
substituted by $B_1$-step). In addition, $A_1$-type configurations
that visit the upper vertex can terminate in the upper triangle,
implying that the last step is of $A_1$ or $A_2$ type, whereas the
step through the right-corner triangle is of type $B$. In
Fig.\ ~\ref{fig:AppA2}(b) it is shown how these configurations can
be obtained from $B$-type configurations. Finally, one obtains
recursion relation for $A_1$-type configurations in the following
form:
\begin{eqnarray}
A_1'&=&\left({p\over 2}
+n_A\right)B^{m-1}B_1^{n+1}A+(p+n_{A1})B^{m}B_1^{n}A_1\nonumber\\
&+&(p+n_{A2})B^{m+1}B_1^{n-1}A_2+n_{C}B^{m} B_1^{n}C\,
.\label{eq:recA1}\end{eqnarray}

\begin{figure*}
\includegraphics[width=130mm]{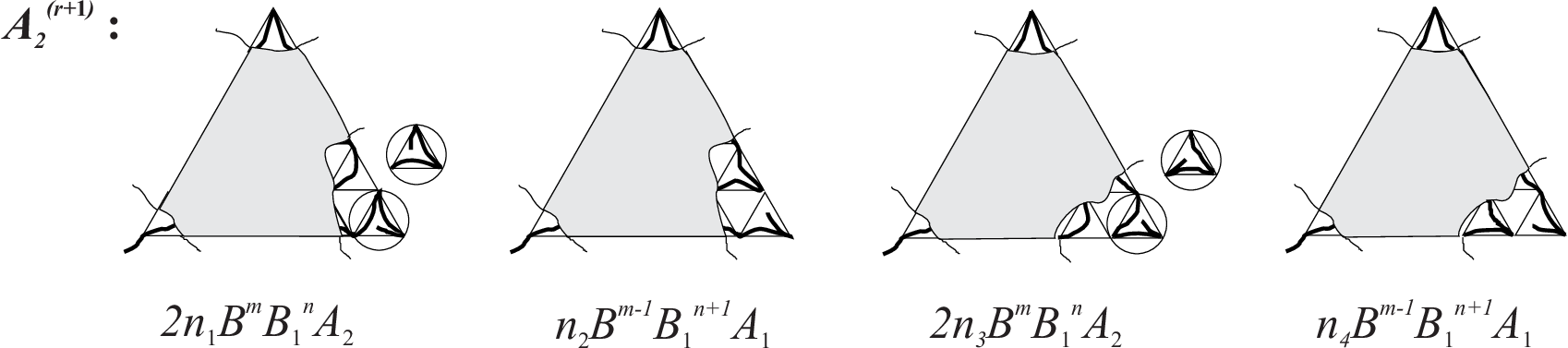}
\caption{Possible configurations of $A_2$-type HWs on $(r+1)$th
order generator of GM fractal, with the ending step in the corner $r$th order generator. \label{fig:AppA3}}\end{figure*}

Since in the case of $A_2$-type configurations the order in which
corner vertices are visited is fixed, corresponding walks from the
class (2) can terminate only in one of the two corner triangles,
lets say in the right corner one.  Then, all possible cases,
obtained from the partitioning of $B_1$-type configurations, are
depicted in Fig.\ ~\ref{fig:AppA3}.
These cases, together with configurations that terminate in interior triangles give the
relation
\begin{eqnarray}
A_2'&=&{{n_A}\over 2}B^{m-2}B_1^{n+2}A+{{p+n_{A1}}\over
2}B^{m-1}B_1^{n+1}A_1\nonumber\\&&\!\!+\left(\frac{n_{A2}}2\!+p\right)B^m\!B_1^n\!A_2\!+{{n_{C}}\over 2}B^{m-\!1}\!
B_1^{n+1}C\, . \label{eq:recA2}\end{eqnarray}
 Finally, with
$a=p+n_A$, $a_1=p+n_{A1}$, $a_2=n_{A2}$, and $c=n_C$ from
(\ref{eq:recA}), (\ref{eq:recA1}), and (\ref{eq:recA2}), one gets
recursion relations (\ref{eq:onelegGM}). Values of the
coefficients $a$, $a_1$, $a_2$, and $c$ for $2\leq b\leq 7$, are
given in Table~\ref{tabela1}.
\begin{table}
\caption{Coefficients $a$, $a_1$, $a_2$, $c$, and $p$, appearing
in recursion relations (\ref{eq:onelegGM}) and (\ref{eq:bezkrake})
for the numbers of $A$, $A_1$, $A_2$, $B$, and $B_1$-type HWs on
GM fractals with $2\leq b \leq 7$, found by computer enumeration
of the corresponding HW configurations.} \label{tabela1}
\begin{ruledtabular}\begin{tabular}{cccccc}
 $b$ & $a$ & $a_1$ & $a_2$& $c$&$p$\\ \hline
  2 & 2 & 2 & 0 & 2 &2\\
  3 & 12& 14 & 4 & 22&8 \\
  4 & 122 &128 & 36 & 212 &40\\
  5 & $1\,842$ & $1\,532$ & 436 & $2\,704$&360 \\
  6 & $35\,390$ & $23\,812$ & $5\,932$ & $42\,368$&3 872 \\
  7 & $880\,646$& $486\,284$ & $110\,876$ & $878\,168$ &62 848\\
\end{tabular}
\end{ruledtabular}
\end{table}
For $b=2$ and $3$ we could find them by direct enumeration of the
corresponding HW configurations, whereas for larger $b$ we had to
use computer facilities.

In a quite similar manner it can be shown that the number of $C$-type
HWs fulfills the recursion relation
\begin{equation}
C'=B^{m}B_1^{n}(a'\frac{B_1}BA+a_1'A_1+a_2'\frac B{B_1}A_2+c'C),
\label{eq:recC}\end{equation}  where coefficients
$a'$, $a_1'$, $a_2'$, and $c'$ depend only on $b$, and their
values for $2\leq b\leq 7$ can be seen in Table~\ref{tabela2}.
\begin{table}
\caption{Coefficients $a'$, $a_1'$, $a_2'$, and $c'$, appearing in
recursion relation (\ref{eq:recC}) for the number of $C$-type HWs
on GM fractals with $2\leq b\leq 7$.} \label{tabela2}
\begin{ruledtabular}\begin{tabular}{ccccc}
 $b$ & $a'$ & $a'_1$ & $a'_2$ & $c'$ \\ \hline
  2 & 1& 2 & 2 & 6 \\
  3 & 12 &16 & 8 & 32\\
  4 & 152 & 168 & 96 & 352 \\
  5 &$2\,544$ & $2\,120$ & 848 &$4\,048$ \\
 6 & $52\,072$ & $35\,152$ &$13\,136$  & $67\,680$  \\
  7 & $1\,340\,536$  & $735\,312$ & $224\,176$  & $1\,374\,944$  \\
\end{tabular}
\end{ruledtabular}
\end{table}

\section{Recursion relations for two-leg HW\lowercase{s} on GM fractals}

Here we prove that the numbers of two-leg HWs, $D$ and $D_1$, on
GM fractals, satisfy recursion relation (\ref{eq:GMTwoLeg}), where coefficients $b_{ij}$ are equal to:
\begin{eqnarray}
 b_{11}&=&  p\,(m+1)B^mB_1^n, \, b_{12}=2p\,nB^{m+1}B_1^{n-1}\, , \nonumber\\
 b_{21}&=&\frac{mp}2 B^{m-1}B_1^{n+1},  \, b_{22}=p\,(n+1)B^mB_1^n \,. \label{eq:DodatakB}
\end{eqnarray}
The two strands which form any two-leg configuration can
terminate either in two different $r$th order generators within
the $(r+1)$th order generator, or in the same one. In the first
case, the corresponding number of walks is of the form
$B^kB_1^lXY$, where $X$ and $Y$ are $A$, $A_1$, $A_2$, or $C$, and $k+l+2=b(b+1)/2$, since all the $r$th order generators must be visited. In the letter case
the number of walks is of the form $B^kB_1^lX$, where $X$ is $D$ or $D_1$, and $k+l+1=b(b+1)/2$. Each such configuration can be
obtained by cutting some $B$ or $B_1$-step of the $B$ or $B_1$
configuration within the $(r+1)$th generator.
\begin{figure*}
\includegraphics[width=130mm]{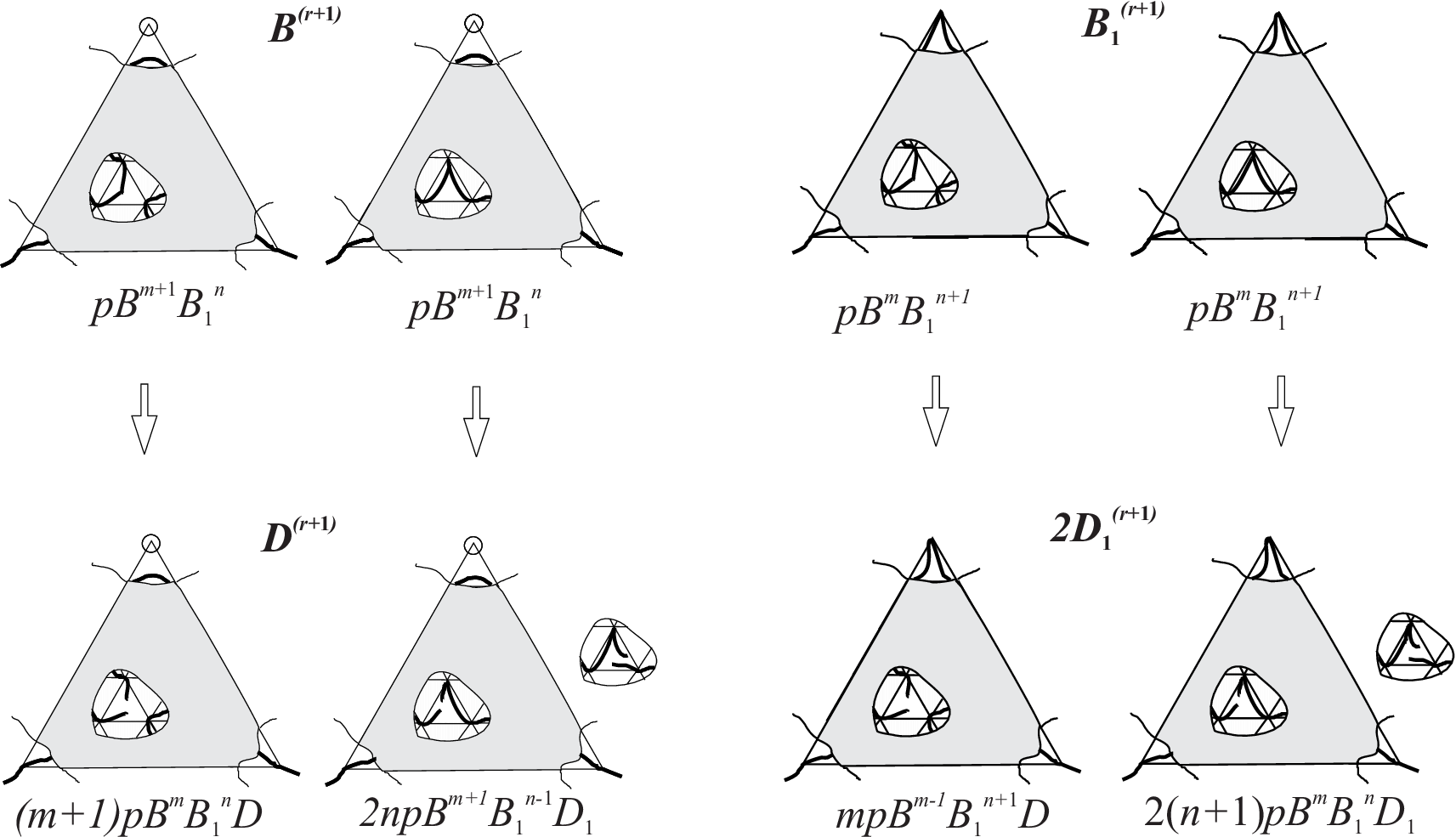}
\caption{All possibilities of cutting $B$ or $B_1$-type
configurations within the $(r+1)$th order generator of GM fractal,
leading to all possible $D$ or $D_1$-type configurations, in which
the two legs terminate in the same $r$th order generator. Terms
under each configuration are equal to the corresponding walks, within the $(r+1)$th order generator. One should notice that by cutting steps of $B_1$ configurations, due to symmetry, doubled number of $D_1$ configurations is obtained (since it is predefined which of the two legs visits the upper corner vertex).}
\label{fig:AppB1}
\end{figure*} All the possibilities
in which this can be done are sketched in Fig.\ ~\ref{fig:AppB1}, so that
(\ref{eq:DodatakB}) directly follows.

\section{Recursion relations for HW\lowercase{s} on
$N$-simplex lattices  \label{ap:nsimplex}}

In this Appendix we give recursion relations for the requisite numbers of HWs,
needed for obtaining the scaling forms for the overall numbers of HWs on $n$-simplex
lattices, for $n=4,5,6$, and 7, together with some relevant details of the derivation of formulas (\ref{eq:scalingClosed}) and (\ref{eq:scalingOpen}).

\subsection*{4-simplex}

Closed HWs on 4-simplex were analyzed in \cite{Bradley}.
Here we quote equations and results relevant for our present
analysis. Recursion relations, as well as the initial values of
the numbers $B_1^{(l)}$ and $B_2^{(l)}$ are
\begin{eqnarray}
B_1'&=&2B_1^4+4B_1^3B_2+6B_1^2B_2^2\, , \quad  B_1^{(1)}=2\,
,\nonumber\\ B_2'&=&B_1^4+4B_1^3B_2+22B_2^4\, , \quad  B_2^{(1)}=1
\end{eqnarray}
The requisite number $x\equiv x_1=B_1/B_2$ satisfies recursion
relation
\begin{equation}
x'=\frac{B_1'}{B_2'}=\frac{2x^4+4x^3+6x^2}{x^4+4x^3+22}\, , \quad
x^{(1)}=2\,. \label{eq:4simx}
\end{equation}
Explicit numerical iteration shows that $x^{(l)}\to 0$, when
$l\to\infty$, so that recursion relation obtains the approximate
form
\[
x^{(l+1)}\approx \frac{3}{11}\left(x^{(l)}\right)^2\, .
\]
This relation implies that $x^{(l)}\sim\lambda_B^{2^l}$, and
precise numerical analysis gives $\lambda_B=0.836620\ldots$. The overall
number of closed HWs on 4-simplex of order $(l+1)$ is equal to
\[
Z_C^{(l+1)}=3\left(B_1^{(l)}\right)^4\, ,
\]
which, with established asymptotic behavior of $x^{(l)}$, for
large $l$ gives the same form as in (\ref{eq:ZatvoreneN}):
\[
Z_C^{(l+1)}\sim \left(\lambda_B^{2^l}B_2^{(l)}\right)^4\, ,
\]
and, consequently, $Z_C^{(l)}\sim \omega^{N_l}(\lambda_B^2)^{N_l^{1/2}}$, with $\omega=1.39911\ldots$.

Directly enumerating one-leg HW configurations, we have found that numbers $A_1$ and $A_2$ satisfy recursion relations of the form (\ref{eq:Jednokrake}) with
\begin{eqnarray} a_{11}&=&6{B_1^2}(B_1+B_2)\, , \quad
a_{21}={B_1^2}(2B_1+3B_2)\, ,\nonumber\\
a_{22}&=&22{B_2^2}(B_1+B_2)+B_1^2(7B_1+16B_2)\, ,\nonumber
\end{eqnarray}
 and $a_{12}=6a_{21}$, so that for parameters $y_1=A_1/B_2$, and
 $y_2=A_2/B_2$, one obtains relation (\ref{eq:YJednokrake}),
 where
\begin{eqnarray}
m_{11}&=&\frac{6x^2\left( 1 + x \right) }{22 + 4x^3 + x^4}  \, , \quad m_{21}=\frac{x^2\left( 3 + 2x \right) }{22 + 4x^3 + x^4}\, ,  \nonumber\\
m_{22}&=&\frac{22 + 22x + 16x^2 + 7x^3}{22 + 4x^3 + x^4}
\, ,\quad m_{12}=6m_{21}\, .\nonumber
\end{eqnarray}
Since $x^{(l)}=B_1^{(l)}/B_2^{(l)}$ tends to 0, it
is obvious that $m_{11}$, $m_{12}$, and $m_{21}$ tend to 0,
whereas $m_{22}\to 1$, when $l\to\infty$, which implies that $y_1$
tends to 0, and $y_2$ to some finite constant value.
Explicit simultaneous iteration of relations (\ref{eq:4simx}) and
(\ref{eq:YJednokrake}), starting with the initial values
$x^{(1)}=2$, $y_1^{(1)}=6$, and $y_2^{(1)}=2$, indeed quickly shows
that $y_1^{(l)}\to 0$, and $y_2^{(l)}\to 62.1081\ldots$.

\begin{figure}
\includegraphics[width=40mm]{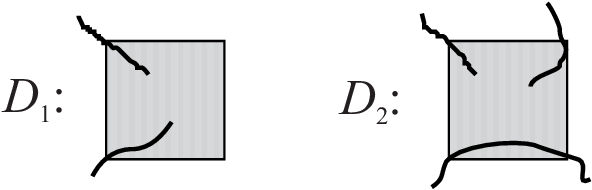}
\caption{Possible types $D_1$ and $D_2$ of two-leg HWs on 4-simplex lattice
of order $l$.}
\label{fig:4simplex}
\end{figure}

The overall number of open HWs on $(l+1)$th order 4-simplex is equal to
\begin{equation}
Z_O^{(l+1)}=12B_1^2\left(A_1^2+2A_1A_2+3A_2^2+B_1D_1\right)\, ,\label{eq:4simplexo}
\end{equation}
where superscript $(l)$ was suppressed on the right-hand side, and $D_1$ is the number of one of two possible two-leg HWs (see Fig.\ ~\ref{fig:4simplex}) on the $l$th order 4-simplex. To obtain asymptotical behavior of the numbers $D_1$ and $D_2$, one needs recursion relations:
\[
\left(
\begin{array}{c}
  D_1^{(l+1)} \\
  D_2^{(l+1)}\\
\end{array}
\right)=\left(
\begin{array}{cc}
  d_{11} & d_{12} \\
  d_{21} & d_{22} \\
\end{array}
\right)\left(
\begin{array}{c}
  D_1^{(l)} \\
  D_2^{(l)}\\
\end{array}
\right)+\left(
\begin{array}{c}
  F_1 \\
  F_2\\
\end{array}
\right)\, , \]
where
\begin{eqnarray}
d_{11}&=&12B_1B_2^2+12B_1^2B_2+8B_1^3,\nonumber\\
d_{21}&=&2B_1^2(B_1+3B_2), \, d_{22}=4B_1^3+88B_2^3,\nonumber
\end{eqnarray}
$d_{12}=4d_{21}$, and $F_1=A_1^2B_1(4B_2+6B_1)+24A_1A_2B_1(B_2+B_1)+A_2^2B_1(42B_1+64B_2)$, $F_2=A_1^2B_1^2+6A_1A_2B_1^2+A_2^2(16B_1^2+44B_1B_2+66B_2^2)$. Introducing new variables $w_1=D_1/B_2$ and $w_2=D_2/B_2$ one gets new recursion relations, which show that $w_1^{(l)}\to 0$ for $l\to\infty$. Since from Eq.~(\ref{eq:4simplexo}) follows that
\[
Z_O^{(l+1)}=12x^2B_2^4(y_1^2+2y_1y_2+3y_2^2+xw_1)\, ,
\]
one finds that two-leg HW configurations do not contribute significantly to the overall number of open HWs for large $l$, and, furthermore, that the following asymptotic formula is valid:
\[
Z_O^{(l+1)}\sim \left(x^{(l)}\right)^2\left(B_2^{(l)}\right)^4\, .
\]
Finally, since $B_2^{(l)}\sim\omega^{4^l}$, and $x^{(l)}\sim \lambda_B^{2^l}$, one obtains $Z_O^{(l)}\sim \omega^{N_l}\lambda_B^{N_l^{1/2}}$.

\subsection*{5-simplex}

According to \cite{Stajic} recursion relations for requisite
numbers $B_1$ and $B_2$, and their initial values, needed for
obtaining the number of closed HWs are such that
$x_1^{(l+1)}\equiv x^{(l+1)}=B_1^{(l+1)}/B_2^{(l+1)}=f_1(x^{(l)})/f(x^{(l)})$, $x^{(1)}=3$, with
\begin{eqnarray}
f_1&=&6 x^5+30 x^4+78x^3+96 x^2+132 x+132,
\nonumber\\
f&=&2 x^5+13 x^4 +32 x^3+88 x^2+220 x+186.\label{eq:5simplexf}
\end{eqnarray}
Numerical iterations give $x^{(l)}\to 0.8023188...$, whereas further analysis of relation (\ref{eq:IteracijaZaOmega}) reveals that $\omega=1.717769...$, and $B_2^{(l)}\sim\omega^{5^l}$.  Since
\[
Z_C^{(l+1)}=12B_1^5 + 30B_1^4 B_2 + 60B_1^3 B_2^2 + 132B_2^5\, ,
\]
it is obvious that for large $l$ one obtains $Z_C^{(l+1)}\sim
\left(B^{(l)}_2\right)^5$, and consequently $Z_C^{(l)}\sim\omega^{5^l}$.

For 5-simplex lattice there are three types of one-leg configurations, $A_1$, $A_2$, and $A_3$, which, as we have found by computer enumeration, fulfil recursion relations (\ref{eq:Jednokrake}), with coefficients $\alpha^{(ij)}_{i_1i_2}$ (see Eq.~(\ref{eq:alfa})) given in Table~\ref{tabela56simplex}.
\begin{table}
\caption{Coefficients $\alpha^{(ij)}$ given by (\ref{eq:alfa}), which appear in the recursion relations (\ref{eq:Jednokrake}), for 5- and 6-simplex lattices.}
\label{tabela56simplex}
\begin{ruledtabular}\begin{tabular}{ccccccccc}
\multicolumn{9}{c}{5-simplex lattice} \\
$i_1$&$i_2$&--&$\alpha^{(11)}_{i_1i_2}$&$\alpha^{(21)}_{i_1i_2}$
&$\alpha^{(22)}_{i_1i_2}$&$\alpha^{(31)}_{i_1i_2}$&$\alpha^{(32)}_{i_1i_2}$
&$\alpha^{(33)}_{i_1i_2}$\\ \hline
0&4&&24&44&548&22&428&1042\\
1&3&&0&44&636&0&296&472\\
2&2&&72&38&374&0&104&0\\
3&1&&72&24&140&4&36&0\\
4&0&&0&6&26&1&6&5\\
\hline
\multicolumn{9}{c}{$\alpha^{(12)}_{i_1i_2}=12\alpha^{(21)}_{i_1i_2}$, $\alpha^{(13)}_{i_1i_2}=24\alpha^{(31)}_{i_1i_2}$, $\alpha^{(23)}_{i_1i_2}=2\alpha^{(32)}_{i_1i_2}$}\\
\hline\hline
\multicolumn{9}{c}{6-simplex lattice}\\
$i_1$&$i_2$&$i_3$&$\alpha^{(11)}_{i_1i_2i_3}$&$\alpha^{(21)}_{i_1i_2i_3}$
&$\alpha^{(22)}_{i_1i_2i_3}$&$\alpha^{(31)}_{i_1i_2i_3}$&$\alpha^{(32)}_{i_1i_2i_3}$
&$\alpha^{(33)}_{i_1i_2i_3}$\\ \hline
0&0&5&0&0&0&0&0&541568 \\
0&1&4&0&0&0&0&0&541568 \\
0&2&3&0&0&57600&0&47168&447232\\
0&3&2&0&0&82848&0&56768 &323264\\
0&4&1&2640&2568&66744&1042&35256&148000\\
0&5&0&2640&1776&23688&614 &10220&33160\\
1 & 2 & 2 & 0 & 0 & 25008 & 0 & 9600 &    47168 \\
1 & 3 & 1 & 0 & 1776 & 45552 & 472 & 17352 & 58464 \\
1 & 4 & 0& 2640 & 2172 & 27768 & 538 & 9576 & 25586 \\
2 & 1 & 2 & 0 & 0 & 4248 & 0 & 0 & 0 \\
2 & 2 & 1 & 0 & 624 & 14616 & 0 & 3460 & 6768 \\
2 & 3 &0& 2280 & 1350 & 15360 & 192 & 3976 & 7708 \\
3&0&2&0&0&456&0&0&0\\
3 & 1 & 1 & 480 & 216 & 2928 & 0 & 480 & 0 \\
3 & 2 & 0 & 1800 & 612 & 5358 & 68 & 1060 & 1188\\
4&0&1&120&36&300&5&56&0\\
4 & 1 & 0 & 720 & 180 & 1182 & 23 & 206 & 168\\
5&0&0&120&24&126&3&20&21\\
\hline
\multicolumn{9}{c}{$\alpha^{(12)}_{i_1i_2i_3}=20\alpha^{(21)}_{i_1i_2i_3}$, $\alpha^{(13)}_{i_1i_2i_3}=120\alpha^{(31)}_{i_1i_2i_3}$, $\alpha^{(23)}_{i_1i_2i_3}=6\alpha^{(32)}_{i_1i_2i_3}$}\\
   \end{tabular}
\end{ruledtabular}
\end{table}
Parameters $y_i^{(l)}=A_i^{(l)}/B_2^{(l)}$ satisfy matrix relation (\ref{eq:YJednokrake}), with all coefficients $m_{ij}$ tending to finite constant values when $l\to\infty$, such that corresponding matrix has one eigenvalue larger than one: $\lambda=5.718...$. This means that $y_i^{(l)}\sim \lambda^l$, which was also confirmed by direct iteration of the recursion relations for $x$ and $y_i$.

Overall number $Z_O^{(l+1)}$ of open HWs is of the form
(\ref{nSimplexOtvorene}), and, similar to the 3-simplex case, it
turns out that terms with the numbers $D_i$ of two-leg HWs have the
same large $l$ behavior as the terms with numbers of one-leg HWs.
Consequently, explicit quoting of recursion relations for $D_i$
numbers is not necessary here, and for the number of open HWs one
obtains: $Z^{(l+1)}_O\sim (B_2^{(l)})^3
A_i^{(l)}A_j^{(l)}\sim\lambda^{2l}(B_2^{(l)})^5$. Then, straightway
follows that $Z_O^{(l)}\sim\omega^{N_l}N_l^\gamma$, with
$\gamma=2\ln\lambda/\ln 5=2.1668...$.

\subsection*{6-simplex}

From the recursion relations for the numbers
$B_1^{(l)}$, $B_2^{(l)}$, and $B_3^{(l)}$, obtained in \cite{Stajic}, certainly follows
that $x_1^{(l+1)}=f_1(x_1^{(l)},x_2^{(l)})/f(x_1^{(l)},x_2^{(l)})$, and
$x_2^{(l+1)}=f_2(x_1^{(l)},x_2^{(l)})/f(x_1^{(l)},x_2^{(l)})$, with $f_1$, $f_2$, and
$f$ being polynomials of order 6. Iterating these relations one concludes that
both $x_1^{(l)}$, and $x_2^{(l)}$ tend to 0, and also $x_2^{(l)}\gg x_1^{(l)}$, so that for large $l$ the following approximate relations are valid:
\[
x_1^{(l+1)}\approx \frac{25008 }{541568}(x_2^{(l)})^4\, , \quad x_2^{(l+1)}\approx
\frac{94336}{541568} (x_2^{(l)})^2\, .
\]
From the latter of these relations follows $ x_2^{(l)}\sim
\lambda_B^{2^l}$, and then, from the first relation one gets $
x_1^{(l)}\sim \left(\lambda_B^2\right)^{2^l}$. Precise numerical
analysis gives $\lambda_B=0.9888...$. Analysis of the corresponding relation (\ref{eq:IteracijaZaOmega}) gives $\omega=2.0550047...$, and $B_3^{(l)}\sim\omega^{6^l}$.
Finally, since $Z_C^{(l+1)}=60 B_1^6 + 360 B_1^5 B_2 + 1170 B_1^4 B_2^2+1920 B_1^3 B_2^3+3960 B_1^2B_2^4 +7920 B_1B_2^5 +5580 B_2^6$, for large $l$ one obtains $Z_C^{(l+1)}\approx \mathrm{const}
\left(B_3^{(l)}x_2^{(l)}\right)^6\sim
\left(\lambda_B^{2^l}B_3^{(l)}\right)^6$, and $Z_C^{(l)}\sim \omega^{6^l}(\lambda_B^3)^{{(6^l)}^{\ln 2/\ln 6}}$.

Numbers $A_1^{(l)}$, $A_2^{(l)}$, and $A_3^{(l)}$ of one-leg HWs for 6-simplex lattice satisfy recursion relations (\ref{eq:Jednokrake}) with coefficients $\alpha^{(ij)}_{i_1i_2i_3}$ given in Table~\ref{tabela56simplex}. Direct numerical iteration of the recursion relations for $x_i$ and $y_i$ parameters, quickly shows that $y_1^{(l)}$ and $y_2^{(l)}$ tend to 0, whereas $y_3^{(l)}$ approaches large, but finite constant: $4.08311... 10^5$. This is in accord with the fact that $m_{33}$ is the only element of the matrix $m$ of the relation (\ref{eq:YJednokrake}) which for large $l$ tends to finite constant, {\em i.e.} not to 0, as do all the other $m_{ij}$. Finally, the overall number (\ref{nSimplexOtvorene}) of open HWs is for large $l$ governed by the term $A_3^2B_2^4$, so that $Z_O^{(l+1)}\sim  \left(x_2^{(l)}\right)^4\left(B_3^{(l)}\right)^6\sim (\lambda_B^{2^l})^4\omega^{6^{l+1}}$,
and $Z_O^{(l)}\sim \omega^{6^l}\left(\lambda_B^2\right)^{(6^l)^{\ln 2/\ln 6}}$.

\subsection*{7-simplex}

To enumerate closed HWs on 7-simplex lattice one needs three
requisite numbers $B_1$, $B_2$ and $B_3$ (see
Fig.\ ~\ref{fig:nsimplexParametri}), which satisfy recursion
relations of the form
\[
B_i^{(l+1)}=\underbrace{\sum\limits_{i_1=0}^{7}\sum\limits_{i_2=0}^{7}\sum\limits_{i_3=0}^{7}}_{i_1+i_2+i_3=7}
b^{(i)}_{i_1i_2i_3}B_1^{i_1}B_2^{i_2}B_3^{i_3}\, , \quad i=1,2,3\, .
\]
 Coefficients $b^{(i)}_{i_1i_2i_3}$, found by explicit computer enumeration of the corresponding HW configurations, are given in
Table~\ref{tabela7simplexB}, whereas initial values are: $B_1^{(1)}=120$, $B_2^{(1)}=24$, and $B_3^{(1)}=3$.  \begin{table}
\caption{Coefficients appearing in recursion relations $B_i'=\sum
b^{(i)}_{i_1i_2i_3}B_1^{i_1}B_2^{i_2}B_3^{i_3}$ for the numbers
$B_i$ ($i=1,2,3$), needed for enumeration of closed HWs on
7-simplex lattice.} \label{tabela7simplexB}
\begin{ruledtabular}\begin{tabular}{cccccc}
 $i_1$&$i_2$&$i_3$&$b^{(1)}_{i_1i_2i_3}$&$b^{(2)}_{i_1i_2i_3}$&$b^{(3)}_{i_1i_2i_3}$\\ \hline
    0 &0 &7&64988160&94599168 &115468800\\
    0 &1 &6 &64988160&141735936&165548544\\
    0 &2 &5 &53667840&118798848&106301952\\
    0 &3 &4 &50112000&78990336 &49499520\\
    0 &4 &3 &32536320&38795520&19747584\\
    0 &5 &2 &14097600&12925632&5819328\\
     0 &6 &1 &3787680& 2712192&1077136\\
      0 &7 &0 &473760&275616& 96864\\
    1&0&6&0&  6498816&11372928\\
     1&1&5&0&10733568& 9748224 \\
     1 &2 &4 &5660160& 15033600&6708480\\
     1 &3 &3 &9319680&13014528&5011200\\
     1 &4 &2 &7985040&7048800&2440224\\
     1 &5 &1 &3459360&2272608&704880\\
      1 &6 &0 &626400&331632&94692\\
      2&1&4&0&566016&0\\
     2 &2 &3 &812160&1397952 &283008\\
    2 &3 &2 &1953600&1597008 &349488\\
    2 &4 &1 &1449360&864840 &199626\\
     2 &5 &0 &396720&187920 &43242\\
     3&1&3&0&54144&0\\
     3 &2 &2 &266880& 195360 &20304\\
     3 &3 &1 &384000&193248 &32560\\
     3 &4 &0 &163800 &66120& 12078\\
      4 &1 &2 &33600&13344 &0\\
      4 &2 &1 &80400&28800  & 3336\\
      4 &3 &0 &50400& 16380 &2400\\
       5 &0 &2 &3720&672 &0\\
      5 &1 &1 &12240&3216 &336\\
      5 &2 &0 &11520 &3024&402\\
      6 &0 &1 &840&204&31\\
       6 &1 &0 &1680&384 & 51\\
       7 &0 &0 & 120 &24&3\\
\end{tabular}
\end{ruledtabular}
\end{table}
Introducing parameters $x_1=B_1/B_3$, $x_2=B_2/B_3$,  one obtains
closed set of recursion relations
\begin{equation}
x_1'=\frac{f_1(x_1,x_2)}{f(x_1,x_2)}\, , \quad
x_2'=\frac{f_2(x_1,x_2)}{f(x_1,x_2)}\,
,\label{eq:7simXrekur}\end{equation} where
\begin{eqnarray}
f_1(x_1,x_2)&=&{\sum\limits_{i_1=0}^{7}\sum\limits_{i_2=0}^{7}
b^{(1)}_{i_1i_2(7-i_1-i_2)}x_1^{i_1}x_2^{i_2}}\, ,\nonumber\\
f_2(x_1,x_2)&=&{\sum\limits_{i_1=0}^{7}\sum\limits_{i_2=0}^{7}
b^{(2)}_{i_1i_2(7-i_1-i_2)}x_1^{i_1}x_2^{i_2}}\, ,\nonumber\\
f(x_1,x_2)&=&{\sum\limits_{i_1=0}^{7}\sum\limits_{i_2=0}^{7}
b^{(3)}_{i_1i_2(7-i_1-i_2)}x_1^{i_1}x_2^{i_2}}\, .\nonumber
\end{eqnarray}
These recursion relations quickly lead to the conclusion that
$x_1^{(l)}\to 0.690015\ldots$, and $x_2^{(l)}\to 1.14695\ldots$,  as
$l\to\infty$. The limiting values are indeed one of the
solutions of the algebraic system of equations
$x_1={f_1(x_1,x_2)}/{f(x_1,x_2)}$, $
x_2={f_2(x_1,x_2)}/{f(x_1,x_2)}$, actually the only positive one.

The overall $Z_C^{(l+1)}$ number of closed HWs on $(l+1)$th order
7-simplex
 is equal to
\[
Z_C^{(l+1)}=\sum\limits_{i_1=0}^{7}\sum\limits_{i_2=0}^{7}c_{ij(7-i-j)}B_1^iB_2^jB_3^{7-i-j}\, ,
\]
which for large $l$ obtains approximate form:
\[
Z_C^{(l+1)}\approx\mathrm{const}\,
\left(B_3^{(l)}\right)^7\, ,
\]
meaning that ${\ln Z_C^{(l+1)}}/{7^{l+1}}\sim {\ln
B_3^{(l)}}/{7^l}$. From the recursion relation for $B_3^{(l)}$ follows equation:
\begin{equation}
\frac{\ln B_3^{(l+1)}}{7^{l+1}}=\frac{\ln
B_3^{(l)}}{7^l}+\frac{1}{7^{l+1}}\ln f(x_1^{(l)},x_2^{(l)})\,
,\label{eq:ZaOmega}
\end{equation}
which can be numerically iterated, together with the recursive
relations (\ref{eq:7simXrekur}) for $x_1$ and $x_2$. In that way,
one finds $\ln\omega=\lim_{l\to\infty}\,\left({\ln
B_3^{(l)}}/{7^l}\right)=0.87382\ldots$,
{\em i.e.} $\omega=2.396056\ldots$. In addition, from relation
(\ref{eq:ZaOmega}) one can write
\[
\frac{\ln B_3^{(k)}}{7^{k}}=\frac{\ln
B_3^{(l)}}{7^{l}}+\sum\limits_{m=l}^{k-1}\frac{1}{7^{m+1}}\ln
f(x_1^{(m)},x_2^{(m)})\, ,
\]
so that when $k\to\infty$ follows
\[
\ln\omega=\frac{\ln
B_3^{(l)}}{7^{l}}+\sum\limits_{m=l}^{\infty}\frac{1}{7^{m+1}}\ln
f(x_1^{(m)},x_2^{(m)})\, .
\]
Since $f(x_1^{(m)},x_2^{(m)})$ decreases with $m$, the above sum
is less then $\frac{\ln f(x_1^{(1)},x_2^{(1)})}{6*7^l}$, and
consequently $7^l\ln\omega>\ln B_3^{(l)}>7^l\ln\omega-5.1$, for all
$l$. This means that for $l\gg 1$ one has $\ln B_3^{(l)}\sim
7^l\ln\omega$, {\em i.e.} $B_3^{(l)}\sim \omega^{7^l}$, finally implying
$Z_C^{(l)}\sim \omega^{N_l}$.

Recursion relations for the numbers $A_i^{(l)}$ of four possible one-leg HWs are too cumbersome to be quoted here, but are available upon request. Numerical analysis of corresponding relations (\ref{eq:YJednokrake}) for parameters $y_i^{(l)}=A_i^{(l)}/B_3^{(l)}$ reveals  that $y_i^{(l)}\sim \lambda^l$, with $\lambda=7.7749...$, being the largest eigenvalue of the matrix $m_{ij}$, obtained in the limit $l\to\infty$. All terms in the formula (\ref{nSimplexOtvorene}) for the number $Z_O^{(l+1)}$ of open HWs are of the same order of magnitude for large $l$, so that
\[
Z_O^{(l+1)}\sim \lambda^{2l}\left(B_3^{(l)}\right)^7  \sim \lambda^{2l} \omega^{7^{l+1}}\, ,
\]
and, consequently:
\[
Z_O^{(l)}\sim \omega^{7^l} \left(7^l\right)^\gamma, \, \gamma=2\frac{\ln\lambda}{\ln 7}=2.10791...\, .\]


\begin{thebibliography}{10}
\bibitem{pocetniuslovi}
It can be shown by induction that for general $n$ the following relations are valid: $B_{i}^{(1)}(n)={(n-i-1)!}/{(i-1)!}$, $A_{i}^{(1)}(n)={(n-i)!}/{(i-1)!}$.
\end{thebibliography}
\end{document}